\documentclass[journal,10pt,twocolumn]{IEEEtran}
\makeatletter
\def\ps@headings{%
\def\@oddhead{\mbox{}\scriptsize\rightmark \hfil \thepage}%
\def\@evenhead{\scriptsize\thepage \hfil \leftmark\mbox{}}%
\def\@oddfoot{}%
\def\@evenfoot{}}
\makeatother
\renewcommand{\baselinestretch}{0.9}
\usepackage{graphicx}
\usepackage{amsmath}
\usepackage{amssymb}
\usepackage{pxfonts}

\newcommand{\gap}{\vspace{2mm}}
\title{SmartConnect: A System for the Design and Deployment of Wireless Sensor Networks}

\author{\IEEEauthorblockN{Abhijit Bhattacharya, Sanjay Motilal Ladwa, 
Rachit Srivastava, Aniruddha Mallya, Akhila Rao,\\ Easwar Vivek. M,
Deeksha G. Rao Sahib, S.V.R. Anand, and Anurag Kumar\\} 
\IEEEauthorblockA{Dept. of Electrical Communication Engineering,
Indian Institute of Science, Bangalore 560012, India.\\
Email: \{abhijit, anand, anurag\}@ece.iisc.ernet.in, \{sanjofpesit, aniruddha.mallya, akhila.suresh.rao, easwar.vivek, deek123, rachitsri\}@gmail.com}}

    \IEEEoverridecommandlockouts

\begin{document}

\maketitle
\vspace{-8mm}
\begin{abstract}
\label{abstract}
We have developed SmartConnect, a tool that addresses the growing need
for the design and deployment of multihop wireless relay networks for
connecting sensors to a control center.  Given the locations of the
sensors, the traffic that each sensor generates, the quality of
service (QoS) requirements, and the potential locations at which
relays can be placed, SmartConnect helps design and deploy a low-cost
wireless multihop relay network. SmartConnect adopts a field
interactive, iterative approach, with model based network design,
field evaluation and relay augmentation performed iteratively until
the desired QoS is met. The design process is based on approximate
combinatorial optimization algorithms. In the paper, we provide the
design choices made in SmartConnect and describe the experimental work
that led to these choices. We provide results from some
experimental deployments. Finally, we conduct an experimental study of the robustness of the network design over long time periods (as channel conditions slowly change), in terms of the relay augmentation and route adaptation required.
\end{abstract}



\section{ Introduction} 
\label{sec:introduction}

Industrial and commercial establishments (such as chemical factories
and hotels) deploy a large number of sensors for control or monitoring
applications. The sensors are typically spread over a large area and
at distances of several tens of meters from the control center.  In
existing installations, the sensors are connected to the control
center by a wireline network, usually a combination of point-to-point
and bus networks. Installation and maintenance of such wireline
networks incur substantial cost. In addition, it is difficult to
expand such wireline sensor networks, for example, to add sensors at
some new locations. Due to such reasons, recently there has been a
spurt of interest in replacing wireline sensor networks with multihop
wireless sensor networks.

There are several sensing applications, particularly in industrial
settings, that could employ low power wireless sensors that use the
wireless physical (PHY) layers and medium access controls (MAC) being
standardized by IEEE 802.15.4 \cite{IEEE}, or Wireless HART
\cite{wireless-hart}, or ISA 100.11a \cite{ISA-100.11a}. Such low
power devices can simply be ``planted'' where needed, and can be
expected to work for several months on batteries and harvested
energy. Due to their low power operation, the range of such radios is
a few meters to a few 10s of meters, necessitating multihopping, and
therefore a higher packet loss rate. There are many applications,
however, e.g., such as data logging and non-critical control (see
\cite{phinney}), for which such low power and lossy networks are
adequate. With such networks in mind, this paper is concerned with the
challenges of designing and deploying wireless relay networks for
interconnecting sensors (viewed as data sources) with a control center
(viewed as a data sink, and also referred to in the paper as a base
station). The system that we have developed to address the challenges,
and the algorithms and procedures embedded in it, is called
\emph{SmartConnect.} In this paper, we present the design of, and
experiences with SmartConnect, a system that iterates by interacting
with partial deployments in the field, and uses on-field measurements
and statistical models, to suggest improvements, eventually leading to
a design that meets QoS requirements.

Given the locations of the sensors and the sink, we are concerned with
the problem of placing wireless relay nodes so that the resulting
multihop wireless network can carry the sensor data to the sink. There
would be placement constraints due to the presence of obstacles (e.g.,
a firewall, a large machine, or a building), or due to taboo regions;
hence we can place relays only at certain designated locations. We
therefore consider the situation in which a number of \emph{potential}
relay locations is provided to the network designer, but as few relays
as possible should be deployed. In addition, since no application can
tolerate arbitrary packet delay and loss, the network design has to
ensure some level of quality of service (QoS). We require that the
network design has to guarantee that the data packets will reach the
control station within a \emph{stipulated delay constraint with a high
  probability}, while taking into account the \emph{highly
  unpredictable nature of wireless channel}. Further, the wireless
network should also preferably have multiple \emph{node disjoint}
paths from each source to the sink to provide \emph{resilience to node
  failures}.

Since there
could be hundreds of locations, a design approach based on an
exhaustive link quality measurement between every possible pair of
locations will be expensive and time consuming. Radio frequency (RF)
propagation models are approximate and cannot yield designs that can
be expected to work when actually deployed. SmartConnect, therefore,
adopts an \emph{iterative field interactive} approach.


The current version of SmartConnect provides a methodology for network
design and deployment for sensor networks that carry low rate
measurement traffic (``light traffic''), typical of applications such
as condition monitoring and non-critical data logging~\cite{phinney}. The methodology comprises the following components:

\noindent
(i) Given the sensor locations, the potential relay locations, and the
location of the sink, \emph{a model} for link quality is used to
generate a graph of potential links over the potential relay locations
(discussed in Section~\ref{sec:link-modeling}).

\noindent
(ii) The QoS constraint for light traffic is formulated in terms of a
Steiner-type problem of minimizing the number of potential relays to
be employed subject to a sensor-sink hop count constraint. This
involves the solution of certain Steiner graph design problems for
which approximation algorithms (developed by us in related prior
work~\cite{iwqos}) are utilized (discussed in
Section~\ref{sec:network-design-approach}).

\noindent
(iii) The proposed relays (typically a very small number, as we found in our experiments)
are placed in the field and link quality measurements are made
under commands from the SmartConnect console. The graph
design algorithm then uses these measured links and models
of the remaining unknown links to propose an improvement
to the design (discussed in Section~\ref{sec:field-interactive-design}, with examples presented in Section~\ref{sec:deployments-results-experiences}).  

\noindent
(iv) A stochastic model from our previous work \cite{winet.srivatasava-etal12performance-analysis} provides an approximate analytical model of multihop networks that use the beaconless CSMA/CA as defined in IEEE 802.15.4, to determine the maximum measurement rate that the design can support while meeting QoS. 

\noindent
(v) At this stage, network operation can start. However, since the quality
of wireless links can vary over time, SmartConnect monitors the
packet delivery performance over the network, and triggers a repair (that may require relay augmentation, or just re-routing) if the performance degrades below a target level (discussed in Section~\ref{sec:robustness}).
 

\section{Related Literature}
\label{sec:related-literature}
Considerable work has been done in the design and deployment of wireless networks in general, and wireless sensor networks, in particular. 
Ray~\cite{ray09planning-analysis-wsn-tool}, Li et al.~\cite{li-etal06planning-deployment-wsn}, and Huang et al.~\cite{huang-etal08deployment-zigbee-wsn} present tools for node deployment to achieve coverage and connectivity in sensor networks, which are based only on modeling and simulation, and do not take into account the unpredictability of wireless links which require on-field testing of modeled links. SmartConnect adopts a field-interactive design approach, where we iteratively improve upon an initial model based design by making on-field link quality measurements. 

Several recent papers address various aspects of wireless link modeling and link quality estimation.

Chipara et al.~\cite{chipara} developed a wall classification based radio coverage prediction model in an indoor WSN, that seems to assume knowledge of the actual path of signal propagation over a link, which is often \emph{not accurately known in a wireless environment due to stochastic fading}. The link model in SmartConnect attempts to capture the average characteristics of the environment by estimating the maximum communication range, $R_{\max}$; deviation in link quality from predicted model due to specific non-homogeneities are accounted for during on-field link learning.

Liu and Cerpa \cite{liu-kerpa} present a three step, \emph{feature
  based} approach to \emph{short temporal} link quality prediction to
better utilize temporally intermediate links for routing
purposes. However, their link prediction approach cannot be adopted in
an iterative \emph{network design} process, since this approach
requires link features (e.g., PRR, RSSI etc.), \emph{which are
  available only for on-field links}, and not for links between
potential locations which are not yet deployed.

Chen and Terzis \cite{chen-terzis} proposed a Bernoulli trial method
to identify spatially intermediate links with high PRR. A key
trade-off of the proposed approach is that one requires several trials
to identify even a single good (high PRR) location. They also
presented a method for \emph{unconstrained} relay placement (i.e., no
restriction on the locations of the relays) to connect a set of
sensing motes to a gateway. Their objective in deploying relays is to
identify \emph{relatively longer} links (beyond the stable connected
range) with high PRR. SmartConnect, on the other hand, addresses a
\emph{constrained} relay placement problem, and provides
\emph{explicit end-to-end QoS guarantee} that cannot be achieved by
just ensuring a high PRR on each link.

Krause et al.~\cite{krause} study the problem of \emph{sensor
  placement} to maximize information obtained from the sensors while
minimizing \emph{total communication cost}. In SmartConnect,
\emph{sensor locations are given}, which is more often the
case~\cite{chen-terzis}; we aim at minimizing the cost of deploying
additional relays, subject to a target communication cost \emph{per}
sensor.

There are also products that deploy relays for sensor connectivity
based only on on-field measurements ~\cite{senzaanalyzer}
~\cite{vykon}. But any broken links are corrected and tested only based on the intuitive prediction of the deployment engineer.

Robinson et al.~\cite{robinson-etal10deploying-nonuniform-propagation}
address the problem of deploying a minimum number of mesh nodes to
build a \emph{tree} for providing client coverage and mesh
(backhaul) connectivity subject to mesh capacity constraints, while
accounting for non-uniform propagation characteristics. A Degree Constrained Terminal Steiner Tree algorithm is used to obtain an initial
design from the estimated network graph. Once deployed, measurements are made \emph{only on the proposed backhaul links} to ensure mesh connectivity. If a predefined SNR threshold is violated, then the network is redesigned
with the refined network graph. 

Beyond the apparent similarity of iterative field measurement driven design, there are several key differences between SmartConnect and the problem addressed by Robinson et al.\cite{robinson-etal10deploying-nonuniform-propagation}.

SmartConnect focuses on providing an end-to-end \emph{QoS guarantee} per source while aiming to minimize the total number of relays, and can provide a \emph{robust design} by allowing for multiple node disjoint, QoS aware paths between each source and the sink. Robinson et al., on the other hand, \emph{do not aim for any explicit end-to-end QoS}, or robustness ($k$-connectivity). Indeed, in their \emph{Measure-and-place} algorithm, measurements are made \emph{only on backhaul links}. Thus, any \emph{poor link quality on a client-mesh link would remain undetected} in this approach, and will affect the end-to-end QoS.

Moreover, once the initial design is deployed on field, SmartConnect
makes measurements among all possible on-field links, thus identifying the potentially good links which were estimated to be bad, and allowing for convergence of the design procedure in a small number of iterations (often just one or two iterations). Robinson et al.\ make measurements only on the \emph{proposed} candidate backhaul links, and not on any other links existing on the field. This keeps the
number of measurements per iteration small, but in turn, may take
several iterations to converge. Also, in their problem, clients cannot
act as mesh nodes, whereas in SmartConnect, sources can act as relays.


\section{Field Interactive Network Design}
\label{sec:field-interactive-design}

\begin{figure}[t]
\begin{center}
\includegraphics[scale=0.35]{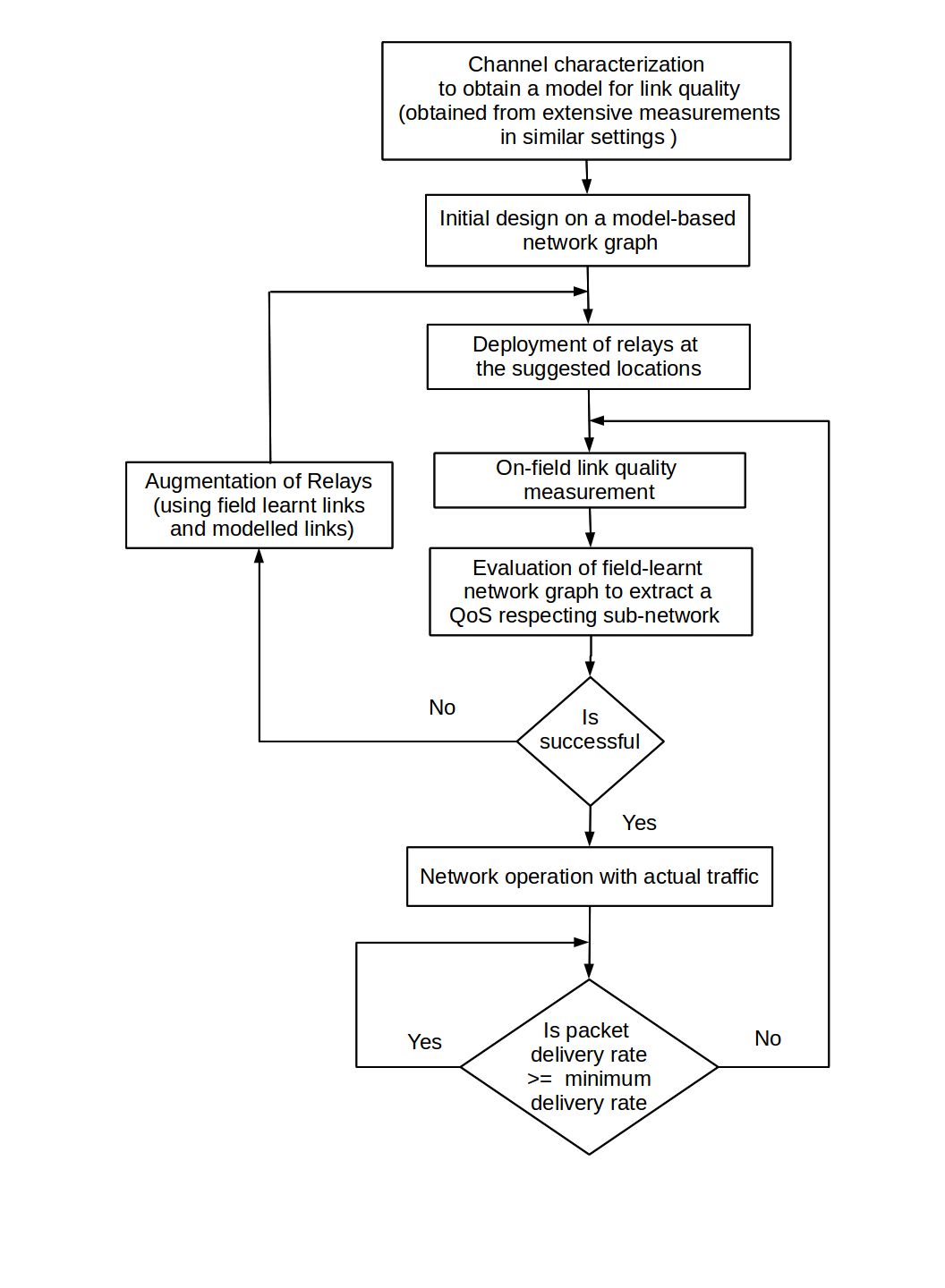}
\end{center}
\vspace{-2mm}
\caption{The phases of SmartConnect's field interactive iterative
  network design, deployment and operation.}
\label{fig:iterative-design-process}
\vspace{-5mm}
\end{figure}

In this section we provide an overview of the network design and
deployment approach utilized by SmartConnect. At the beginning of the
design process we are given a deployment region with designated sensor
locations, the location of the sink, and several potential locations
at which wireless relays can be placed.

To design a network connecting the set of sensor sources and the sink,
using topology design algorithms, we need a network graph defined over
the sensor sources, the potential relay locations and the sink. In
very small networks, relays can be placed at all potential relay
locations and the qualities of the links between every pair of relay
locations could be learnt by on-field measurements. Using our
previously developed topology design algorithm \cite{iwqos} on such a
network graph of field-learnt links would provide a one-shot design
satisfying QoS constraints, if such a design is feasible. For a larger
network with a large number of potential relay locations, deploying
relay nodes at every potential relay location would be
impractical. What we need is a model to capture the characteristics of
the wireless channel in the deployment environment so as to predict
feasible links between the locations of the sources, the relay
locations, and the sink; we can then apply the design algorithm on the
model based network graph to obtain an initial design satisfying QoS
constraints on this graph, and place relays only at the locations
suggested by this \emph{initial design}.
     
While there could be several approaches for modeling the quality of
links (e.g., an RF propagation modeling tool could be utilized), we
have adopted a simple link quality model. Any two nodes within a
distance of $R_{\mathsf{max}}$~meters are predicted to be in
communication range of each other (details of the procedure to obtain
such a link model are provided in
Section~\ref{sec:link-modeling}). This link model, i.e., $R_{\max}$,
can be significantly different for different deployment environments
such as an outdoor power distribution yard, or an indoor industrial or
commercial establishment, etc. We note that $R_{\max}$ is just a
simple distillation of statistical data collected in a similar
environment, and merely asserts that links shorter than $R_{\max}$ are
likely to be good (in a sense to be explained in detail in Section~\ref{sec:link-modeling}).

Statistical link models can only estimate the channel characteristics
of the environment, but cannot fully ascertain the existence of the
predicted links on the field. Some links could be worse than predicted
due to the presence of large obstacles, or even better than predicted,
e.g., due to line of sight visibility. \emph{Actual on-field link
  quality measurement is, therefore, needed} before the network can be
put into operation. By actually placing relay nodes at locations
provided by the initial design, we can learn the \emph{on-field link
  quality} of all the links between the deployed nodes. Upon completion of
link learning, we have a network graph of \emph{field-learnt links}
(acceptable links). This graph is fed to the topology design algorithm
for \emph{evaluation} to verify whether the on-field nodes along with
the good quality learnt links are sufficient to obtain a sub-network
connecting the sensors to the sink, while meeting QoS. If this
evaluation of the field-learnt network graph is successful, then the
design is complete. Only the relays that are part of this topology are kept, and the rest are removed. 

If however, the topology design algorithm cannot extract any QoS
respecting subnetwork from the on-field network graph on the deployed
nodes, the network will need \emph{relay augmentation}. At this stage
we have \emph{learnt links} between the nodes on field, and
\emph{modeled links} between the rest of the locations. This network
graph consisting of modeled and learnt links is now used by the
topology design algorithm to obtain a subnetwork that meets the
required QoS, which will require the deployment of relays at
additional potential locations. Since the locations suggested for
relay augmentation are based on modeled links, the newly added links
(due to relay augmentation) need to be learnt on field before
re-evaluation. As shown in Figure~\ref{fig:iterative-design-process},
after the initial design, \emph{link learning}, \emph{evaluation} and
\emph{augmentation} are repeated iteratively until a QoS respecting
network is obtained.  This is the crux of SmartConnect's \emph{field
  interactive, iterative design}. The iterative process provides a
method of partial deployment of networks with modeled and learnt links
until a complete deployment meeting QoS requirements on field is
obtained. Provided that the link model is not too conservative (i.e.,
it does not severely underestimate the link quality), if there exists
a QoS respecting subnetwork in the actual on-field network graph, this
iterative procedure will converge to such a solution on field after a
few iterations (a possible problem with an overly conservative link
model could be that the topology design algorithm may declare the
problem to be QoS-infeasible even if there exists a feasible solution
in the actual on-field network graph). A remedy for this situation is
discussed in Section~\ref{sec:practical-issues}.

Additionally, since
wireless links are highly dynamic, with significant changes in time of
day and surrounding activity, robustness of a deployed network is a
challenging issue. To account for these link variations we perform a
continuous repair process. As we receive sensor data from the sensor
sources, knowing the rate at which data is being sent, we measure its
packet delivery rate. This packet delivery rate is continuously
monitored at the base station. If it reduces below the delivery rate
the network was designed for, then a repair is triggered from the base
station. Link learning is initiated,
the updated network graph is evaluated, and augmentation is performed
if necessary. On arriving at a design that was successfully evaluated,
network operation continues till repair is triggered again. This
procedure makes the network robust by accounting for link variation
over large time duration by repairing as and when necessary. Link
variations that stem from change in activity in the environment,
change in obstacle profile or seasonal changes can be handled. 

The details and implementation of the concepts presented in the current
section are discussed in the rest of the paper. For a virtual demonstration of SmartConnect, see \cite{sc-demo}. 
\section{Wireless Link Modeling}
\label{sec:link-modeling}

\begin{figure}
\includegraphics[scale=0.26]{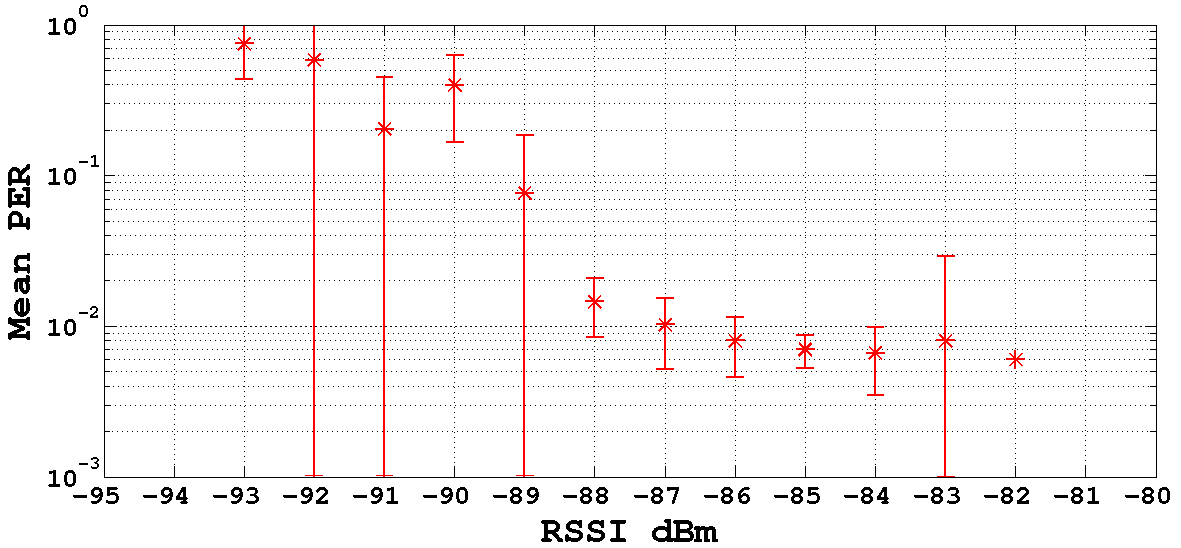}
\caption{PER vs RSSI measured between two motes, connected by a standard coaxial cable and standard attenuators. ``Over-the-air'' packet size: 120 bytes}
\label{plot:PERvsRSSI-attenuators}
\vspace{-5mm}
\end{figure}

We assume that our network carries packets of size 120 bytes (in
particular, the packet consists of 90 bytes of payload, together with 24 bytes MAC header, and 6 bytes PHY header); our measure of link quality is in terms of
packet error rate (PER). The packet error rate is determined by the
bit error rate, which in turn is governed by the received signal
strength, the noise and the interference, and the
modulation-demodulation scheme. For the particular radios we have used,
Figure~\ref{plot:PERvsRSSI-attenuators} shows measured PER (for our
standard packet sizes) versus the RSSI (received signal strength
indicated by the receiver) in a controlled experiment (see
\cite{raman-etal06link-(in)stability},
\cite{srinivasan-etal06RSSI-underappreciated}). The PER measurement
was conducted by connecting two TelosB motes back-to-back via
standard attenuators, and varying the RSSI value. The experiment was
repeated for several different node pairs, and the mean PER over all
the experiments was obtained as a function of
RSSI. Figure~\ref{plot:PERvsRSSI-attenuators} shows the mean PER as
well as the 95\% confidence interval as a function of RSSI. We notice
that the PER is reliably below 0.02 or 0.03 for RSSI values larger
than -88~dBm, whereas below this RSSI value not only does the PER
rapidly increase, but is highly variable \emph{from mote to mote}. We
conclude from this experiment that an on-field link should have an
RSSI of better than $-88$~dBm.

Given the above experimental results, for our iterative design process
we seek a simple link model, in terms of a link length $R_{\max}$ such
that with a transmitter power of 0~dBm, a receiver at a distance of
$\leq R_{\max}$ is very likely to receive a signal strength better
than -88~dBm.  The transmitter power of 0~dBm \footnote{The highest power level in this device.} is chosen so as to minimize the requirement of relays.

We now present our approach for choosing $R_{\max}$. In this process we have to contend with wireless propagation, a highly unpredictable phenomenon. Classically, for the purpose of analyzing wireless digital communication links, the RF propagation loss is modeled in terms of (i) a nominal path loss model (typically an inverse power law model), (ii) a stochastic shadowing model (which accounts for statistical variation of path loss over different links of the same length), and (iii) a stochastic fading model (which accounts for multipath fading and channel variations). 

In choosing $R_{\max}$, we define three measures:

\begin{description}

\item[$q_{\max}$] The maximum target PER (e.g., 0.05; see the
  measurement results in Figure~\ref{plot:PERvsRSSI-attenuators});
  equivalently we can think in terms of the minimum RSSI,
  $\mathrm{RSSI}_{\min}$, e.g., $-88$~dBm.

\item[$p_{\mathrm{out}}$] The fraction of time that the PER on the
  link is worse than $q_{\max}$; since links do fade over time, outage
  is inevitable; the probability of a multihop path being in outage
  increases with the number of hops; hence, we need to have a target link outage probability (henceforth denoted by $P_{\mathrm{out}}$).

\item[$p_{\mathrm{bad}}$] is a function of the link length  $R$, and 
is defined as the fraction of links of length $R$ that do not meet 
the outage target $P_{\mathrm{out}}$; this measure is relevant since,
  due to shadowing, there are link to link average path loss
  variations, even for links of a given length.
\end{description}

\noindent
We also have a target $p_{\mathrm{bad}}$, which we call $P_{\mathrm{bad}}$. 
The consequences of the choices of $P_{\mathrm{out}}$ and $P_{\mathrm{bad}}$, 
and a methodology for making these choices will be presented below. 

Having defined these measures and their targets, we then define
\begin{description}
\item[$R_{\max}$] The link length $R$ at which $p_{\mathrm{bad}}$ 
 is less than or equal to $P_{\mathrm{bad}}$.
\end{description}

Once we identify an $R_{\max}$ such that targets for all the above
measures are met, then in the design steps that involve model based
design (see Section~\ref{sec:field-interactive-design}), we just
include all links that are of length $\leq R_{\max}$. In doing this,
the measure $p_{\mathrm{bad}}$ plays two roles: (i) The larger the
value of $P_{\mathrm{bad}}$, the larger the probability that the
model-based design will not meet QoS on the field. (ii) It also helps
to determine the set of potential locations, as follows. Given $R_{\max}$ and
$P_{\mathrm{bad}}$, the set of potential locations can be chosen to be
such that if we consider the graph on these locations with all edges
of length $\leq R_{\max}$, and if each such potential edge is removed
with probability $P_{\mathrm{bad}}$, then with a high probability the
remaining graph still has a subgraph that meets our QoS objectives.

\begin{figure}[t]
\centering
\includegraphics[scale=0.25]{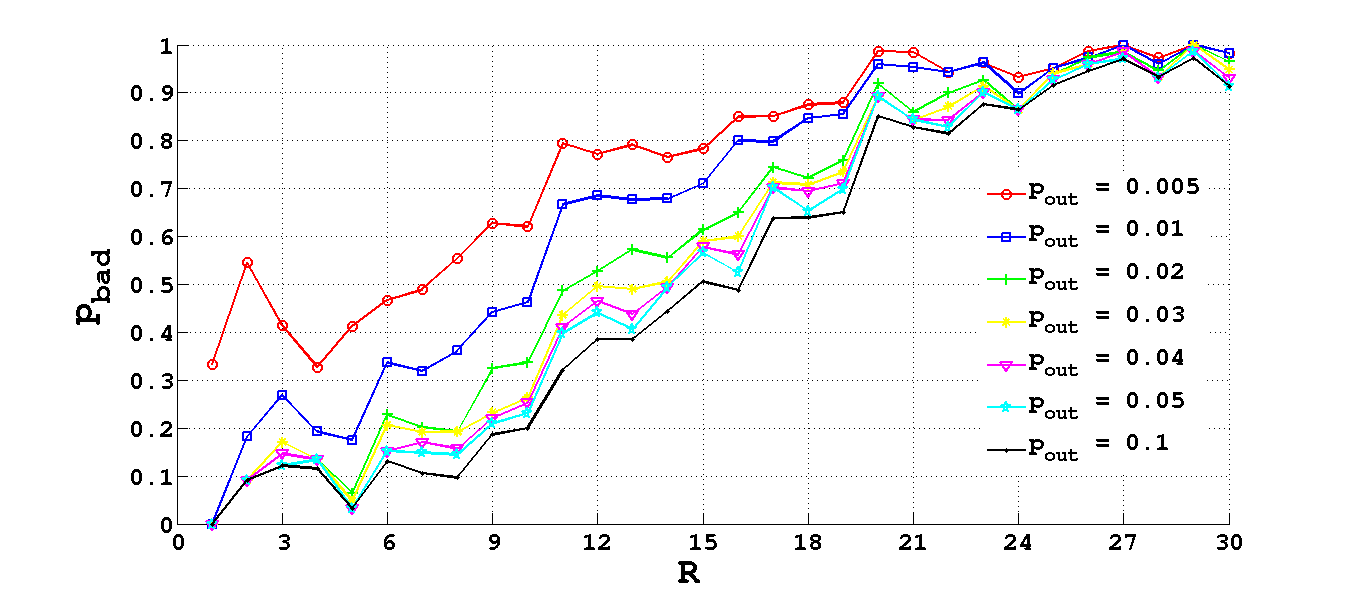}
\caption{Measurements taken inside our department building:
  $p_{\mathrm{bad}}$, i.e., the fraction of links whose outage
  probability is greater than $P_{\mathsf{out}}$, vs link length, $R$,
  plotted for multiple values of $P_{\mathsf{out}}$.}
\label{plot:building-pout-graph}
\vspace{-5mm}
\end{figure}

Although analytical models (e.g., Rayleigh or Ricean fading, and
log-normal shadowing) can be used to relate $R_{\max}$,
$\mathrm{RSSI}_{\min}$, $P_{\mathrm{out}}$, and $P_{\mathrm{bad}}$,
these relations are only indicative and cannot be used for reliably
characterizing the quality of links in a design process (particularly, in a nonhomogeneous setting such as the interiors of a building). We,
therefore, resort to a measurement-based approach.
\begin{itemize}
\item A large number of nodes (50 nodes in our experiments) are scattered throughout the region, so as to obtain links of varied distances.
\item Each node, one after the other, broadcasts a large number (5000
  packets in our experiments) of ``hello'' packets. The nodes that
  receive them, log the received signal strength (RSSI) of each packet
  along with the details of the sender node.
\item We now have a distribution of RSSI for every link in the
  network. These distributions are distilled into the plots shown in
  Figure~\ref{plot:p_bad_vs_r} and
  Figure~\ref{plot:building-pout-graph}, as explained in the following
  bullet points.
\item From the graph in Figure~\ref{plot:PERvsRSSI-attenuators} we see
  that to obtain a PER of less than or equal to 0.05 with high
  probability, the RSSI along the link should be greater than or equal
  to $-88$~dBm; this is when the link is not in outage.
\item The probability that a link is not in outage, say,
  $p_{\mathsf{nout}}$, is the fraction of packets received at RSSI
  $\geq -88$~dBm on the link. The probability that the link is in
  outage, $p_{\mathsf{out}}$, is estimated as $ 1- p_{\mathsf{nout}}$.
\item A link is said to be `good' if its outage probability is less
  than $P_{\mathsf{out}}$, else it is termed `bad'.
\item The links we have are now binned according to
  their link length. The link length is rounded off to the nearest meter 
  to make one meter bins. In each bin, we compute the fraction, say,
  $p_{\mathsf{good}}$, of `good' links, and then $p_{\mathsf{bad}}$ is
  estimated as $1-p_{\mathsf{good}}$.
\item \emph{The maximum distance bin in which the probability of a link
  being bad ($p_{\mathsf{bad}}$) is less than or equal to a threshold
  $P_{\mathsf{bad}}$
  is chosen as the maximum communication range ($R_{\max}$) for reliable
  communication.}
\item The choice of $P_{\mathsf{out}}$ and
  $P_{\mathsf{bad}}$ are based on measurements and is elaborated in the two
  examples below. 
\end{itemize}

\begin{figure}[t] 
\begin{center} 
\includegraphics[scale =0.046]{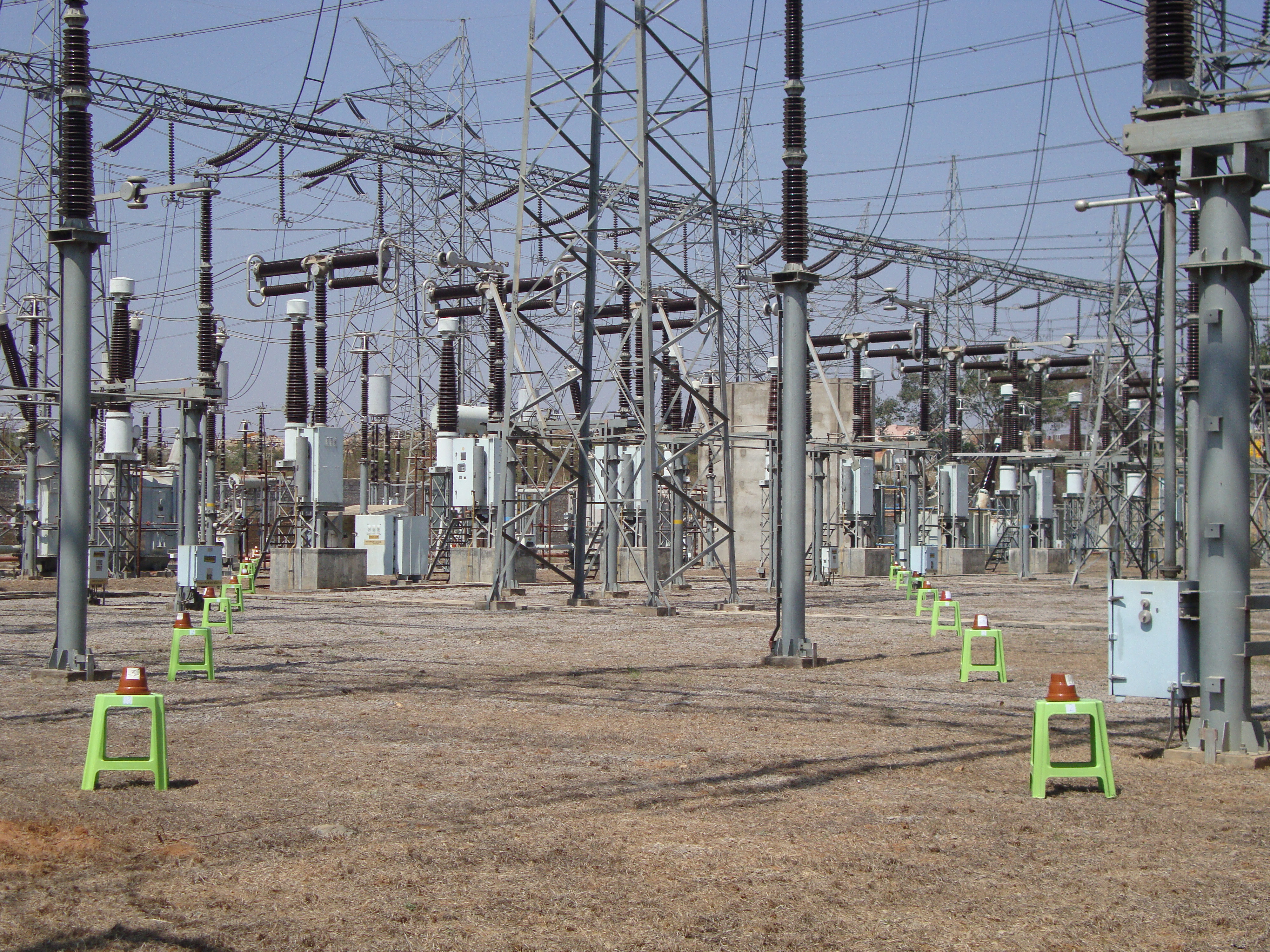} 
\end{center}
\caption{A view of the 440 KV power distribution yard, showing the layout of equipment; apart from the transmission towers, several transformers and a firewall can be seen at a distance.}
\label{fig:nelamangala-location} 
\vspace{-2mm}
\end{figure}

\begin{figure}[t]
\centering
\includegraphics[scale=0.25]{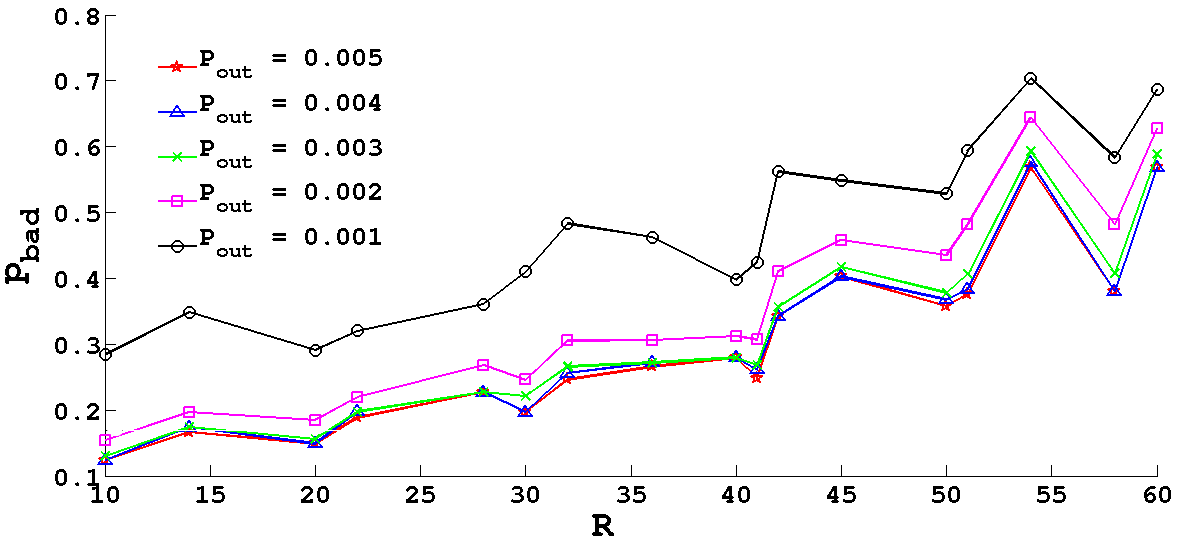}
\caption{Measurements taken from a power distribution yard environment
  (see Figure~\ref{fig:nelamangala-location}): $p_{\mathrm{bad}}$,
  i.e., the fraction of links whose outage probability is greater than
  $P_{\mathsf{out}}$, vs link length, $R$, plotted for multiple values
  of $P_{\mathsf{out}}$.}
\label{plot:p_bad_vs_r}
\vspace{-5mm}
\end{figure}

We have carried out such measurements in a couple of different environments: a 440KV outdoor power distribution yard (since the goal of our project was to create wireless networks for connecting sensors in such yards), and our department building (ECE Department, IISc; a layout diagram is shown in Figure~\ref{fig:building_design_phase}).

In Figure~\ref{plot:p_bad_vs_r}, we provide a summary of measurements
that we took at a 440KV outdoor power distribution yard. A photograph
of a part of this yard is shown in
Figure~\ref{fig:nelamangala-location}. There are several tall towers
across which are strung high-tension power cables; there are
transformers, circuit breakers, and firewalls separating the
transformer bays. The ground is covered with coarse gravel; there
are also drainage ditches, and narrow tarred roads criss-crossing the
area.  

In Figure~\ref{plot:p_bad_vs_r}, for each value of $R$, between 10~m
and 60~m (on the $\mathsf{x}$-axis), we show (on the
$\mathsf{y}$-axis) $p_{\mathrm{bad}}$, the fraction of links whose
outage probability was worse than each of five values of
$P_{\mathrm{out}}$ (0.001, 0.002, 0.003, 0.004, 0.005). In these
measurements, the target $RSSI$ was -88~dBm. As expected, for a fixed
value of $P_{\mathrm{out}}$, $p_{\mathrm{bad}}$ increases with
$R$. For example, with $P_{\mathrm{out}} = 0.002$, for $R=30$m, about
25\% of the links displayed an outage probability worse than 0.002,
whereas with $R=60$m this went up to more than 60\%. We notice that
there is a sharp increase in $p_{\mathrm{bad}}$, for every value of
$R$, if $P_{\mathrm{out}}$ exceeds 0.002.  From the plot we
also see that, at every distance, at least 10\% of the links are
bad. Lowering $P_{\mathsf{out}}$ will give a more conservative value
of $R_{\mathsf{max}}$, for a chosen $P_{\mathrm{bad}}$.
 So the choice of $R_{\mathsf{max}}$ is a trade-off between 
$P_{\mathsf{out}}$ and $P_{\mathsf{bad}}$. For a given 
$P_{\mathsf{bad}}$, increasing $P_{\mathsf{out}}$ may increase 
$R_{\max}$, but
affects the packet delivery probability, whereas reducing 
$P_{\mathsf{out}}$ reduces $R_{\max}$, thus leading to a 
more conservative (and possibly costly) design. For a given 
$P_{\mathsf{out}}$, increasing $P_{\mathsf{bad}}$ can increase 
$R_{\max}$, but it also increases the chance that a proposed 
design requires augmentation, and requires more potential relay 
locations to begin with. Based on the measurement
results shown in Figure~\ref{plot:p_bad_vs_r}, we chose
$P_{\mathrm{out}} = 0.004$, $P_{\mathrm{bad}} = 20$\%, yielding 
$R_{\max} = 30$m.

Figure~\ref{plot:building-pout-graph} shows the summary of similar
measurements we made for an indoor deployment inside our department
building. The analysis of the figure, as in the previous example,
tells us that for a $P_{\mathsf{bad}}$ of less than 10\% we get
a value of $R_{\mathsf{max}}$ of only $3$ meters, for any value of
link outage. We see that the link outage in the indoor case is much
larger than that of the outdoor power distribution yard case. So based
on Figure~\ref{plot:building-pout-graph}, we chose $P_{\mathsf{out}} =
0.04$, $P_{\mathrm{bad}} = 20$\%, yielding $R_{\mathsf{max}} = 8$m.


\section{Network Design Approach}
\label{sec:network-design-approach}

\vspace{2mm} In each iteration of the design process outlined in
Section~\ref{sec:field-interactive-design}, we have a graph on the set
of sources and potential relay locations. In some steps, the graph is
based only on the model discussed in Section~\ref{sec:link-modeling},
i.e., all pairs of nodes separated by less than $R_{\max}$ meters are
assumed to have a good link between them, or based on measurements, or
both. In \cite{iwqos} and \cite{bhattacharya-techreport}, we have
elaborated how, given a network graph defined on the source nodes, the
potential relay locations, and the Base Station (BS), a candidate
topology satisfying the QoS constraints is extracted. The basic
network design problem that we want to address can be stated as
follows: Given a network graph $G = (V, E)$, where $V = Q\cup P$, is
the set of vertices consisting of source nodes $Q$ (including the base
station) and potential relay locations $P$, and $E$ is the set of all
feasible links, obtain a subnetwork that connects the source nodes to
the base station with the requirement that
\begin{enumerate}
\item A minimum number of relay nodes is used.
\item There are at least $k$ node disjoint paths from each source node to the BS.
\item The maximum delay on any path is bounded by a given value $d_{\max}$, and the packet delivery probability (the probability of delivering a packet within the delay bound) on any path is $\geq p_{\mathrm{del}}$.
\end{enumerate}

The following assumptions are made regarding the network traffic, and the nature of the wireless medium.

\begin{enumerate}
\item The traffic generated by the sensor nodes is \emph{very} light;
  so there is \emph{rarely} more than one packet in the network at any
  point of time so that, with a high probability, the network is
  contention free. Such a situation can arise in many applications
  where successive measurements being taken are well separated in time
  so that the measurements can be ``staggered'', and they do not
  occupy the medium at the same time, e.g., applications such as data
  logging, and non-critical preventive control (see
  \cite[p.~9]{phinney}).
\item As mentioned in Section~\ref{sec:link-modeling}, since there is
  a non-zero PER on each link, packet losses due to random channel
  errors have been considered, so that a random number of
  retransmissions are required until each packet is delivered across
  each link, or is dropped due to excessive retransmissions.
\item Also \emph{slow fading} is permitted so that the packet error
  probabilities on the links vary slowly over time, leading to
  possible \emph{link outage} (See Section~\ref{sec:link-modeling}).
\end{enumerate}

\emph{We approach the problem by designing the network for the, so
  called, ``lone packet model'', thus reducing the problem to one of
  graph design, and then using an analytical model to evaluate the
  maximum data rate that the network can support while meeting QoS.}
Also note that in order to meet the QoS constraints for a positive
traffic arrival rate, it is \emph{necessary} to satisfy the QoS
constraints under the lone-packet model \cite{bhattacharya-techreport}. As it turns out, even this simplified version of the problem is NP-Hard. Therefore, one cannot hope to solve the more complex general problem with positive arrival
rate unless one has a satisfactory solution to this basic lone-packet
design problem.

We outline below, how, under a lone-packet model, we can reduce the
QoS constrained network design problem into a graph design problem.

\subsection{Mapping of QoS to Hop Constraint}
\label{sec:qos-to-hop-constraint}
Under the assumptions stated earlier, we present below, an elementary
analysis that maps the QoS constraints, namely, maximum end-to-end packet
delay, $d_{\max}$, and packet delivery probability, $p_{\mathrm{del}}$, to a \emph{hop count bound} $h_{\max}$ on each path from each source to the BS. 

Before proceeding further, we summarize for our convenience, the notation used in the development of the model.

\vspace{2mm}
\textbf{User requirements:}
\begin{description}
\item[$L$] The longest distance from a source to the base-station (in meters)
\item[$k$] The required number of node disjoint paths between each source and
  the base-station
\item[${d}_{\max}$] The maximum acceptable end-to-end delay of a
  packet sent by a source (packet length is assumed to be fixed and
  given)
\item[$p_{\mathrm{del}}$] Packet delivery probability: the probability
  that a packet is not dropped \emph{and} meets the delay bound
  (assuming that at least one path is available from each source to
  the base station).
\end{description}
\vspace{2mm}
\textbf{Parameters obtained from the standard:}
\begin{description}
\item[$D_q(\cdot)$] The cumulative distribution function of packet
  delay on a link with PER $q$, given that the packet is not dropped;
  $D^{(h)}_q(\cdot)$ denotes the $h$-fold convolution of
  $D_q(\cdot)$. Under the lone packet model, $D_q(\cdot)$ is obtained
  by a simple analysis of the backoff and attempt process at a node,
  as defined in the IEEE~802.15.4 standard for beaconless mesh
  networks. 
\item[$b(\cdot)$] The mapping from SNR to link BER for the modulation scheme (see \cite{abhijitmeth})
\item[$\delta(\cdot)$] The mapping from PER to packet drop probability
  over a link (see \cite{abhijitmeth}). Note that even when there is no contention,
  packets could be lost due to random channel errors on links (i.e.,
  non-zero link PER). A failed packet transmission is reattempted at
  most three times before being dropped.
\end{description}
\vspace{2mm}
\textbf{Design parameters:}
\begin{description}
\item[$P_{\mathrm{xmt}}$] The transmit power over a link (assumed here
  to be the same for all nodes)
\item[$\gamma_{\min}$] The target SNR on a link
\item[$q_{\max}$] The target maximum PER on a link, when not in outage
\end{description}
\vspace{2mm}
\textbf{Parameters obtained by making field measurements:}
\begin{description}
\item[$R_{\max}$] The maximum allowed length of a link on the field to meet the target SNR, and outage probability requirements
\item[$P_{\mathrm{out}}$] The maximum probability of a link SNR falling below
  $\gamma_{\min}$ due to temporal variations. A link is ``bad'' if its
  outage probability is worse than $P_{\mathrm{out}}$, and ``good'', otherwise
\end{description}

\vspace{2mm}
\textbf{To be derived:}
\begin{description}
\item [$h_{\max}$] The hop count bound on each path, required to meet the packet delivery objectives
\end{description} 

\noindent
\textbf{Remark:} In practice, $k$, the number of node disjoint paths from each source to the sink, can be chosen so that a network monitoring and repair process ensures that a path is available from each source to the BS at all times. The choice of $k$ is not in the scope of our current formulation, and would depend on the rate at which paths can fail, how quickly the network monitoring process can detect node failures, and how rapidly the network can be repaired. We, thus, assume that, whenever a packet needs to be delivered from  a source to the BS, there is a path available, and, by appropriate choice of the path parameters (the length of each link, and the number of hops), we ensure the delivery probability, $p_{\mathrm{del}}$.

\subsubsection{Design constraints from packet delivery objectives}

Consider, in the final design, a path between a source $i$ and the
base-station, which is $L_i$ meters away.  Suppose that this path has
$h_i$ hops, and the length of the $j$th hop on this path is $r_{i,j},
1 \leq j \leq h_i$. Then we can write
\begin{eqnarray}
\label{eqn:hopcount-bound-from-rmax}
  L_i \leq \sum_{j=1}^{h_i} r_{i,j} \leq h_i R_{\max}
\end{eqnarray}
where the first inequality derives from the triangle inequality, and
the second inequality is obvious. Since $L$ is the farthest that any
source is from the base station, we can conclude that the number of
hops on any path from a source to a sink is bounded below by
$\frac{L}{R_{\max}}$.

Suppose that we have obtained a network in which there are $k$ node
independent paths from each source to the base-station, and all the
links on these paths are good (``good'' in the sense explained earlier in the definition of $p_{\mathrm{out}}$). Following a conservative approach, we take the
PER (conditioned on the link not being in outage) on every good link to be $q_{\max}$ (we are taking the worst case PER on each link, and are not accounting for a lower PER on a shorter link).
Consider a packet arriving at Source~$i$, for which, by design, there are $k$
paths, with hop counts $h_\ell, 1 \leq \ell \leq k$, and suppose that
at least one of these paths is available (i.e., all the nodes along
that path are functioning). The availability of such a path is determined by a separate route management algorithm that monitors the $k$ routes for each source, and if the currently active route from any source fails to provide the target QoS (e.g., delivery probability), selects one of the remaining good paths to route traffic from that source to the sink. The path selection algorithm would incorporate a load and energy balancing strategy. 

If the chosen path has $h_{\ell}$ hops in it, then the probability that none of the edges along the chosen path is in outage is given by
\begin{eqnarray*}
 (1 - P_{\mathrm{out}})^{h_\ell}
\end{eqnarray*}
Increasing $h_{\ell}$ makes this probability smaller. With this in
mind, let us seek an $h_{\max}$, by the following conservative
approach. First, we lower bound the probability of the chosen path not
being in outage by
\begin{eqnarray*}
   (1 - P_{\mathrm{out}})^{h_{\max}}
\end{eqnarray*}
Now (recalling the definitions earlier in this section) we can ensure that the packet delivery constraint is met by requiring
\begin{eqnarray}
  \label{eqn:using-the-pdel-constraint}
  (1 - P_{\mathrm{out}})^{h_{\max}}(1 - \delta(q_{\max}))^{h_{\max}} D^{(h_{\max})}_{q_{\max}} (d_{\max})\geq p_{\mathrm{del}}
\end{eqnarray}
where, the second term lower bounds the packet success probability on a path, given the path is not in outage, and the last term is the
probability of \emph{in-time} delivery given that the packet was not
dropped. Recall that we take the PER on each ``good'' link (when not in outage) to be $q_{\max}$.

Since the packet drop probability, given the path is not in outage, is
negligibly small (in fact, it is upper bounded by $q_{\max}^n$ on each
link, where $n$ is the total number of failed transmission attempts
before a packet is dropped; $n=4$ for IEEE~802.15.4 CSMA/CA MAC
\cite{IEEE}), the packet delivery probability is essentially dominated
by the path outage probability, and the probability of \emph{in-time}
delivery, given the packet is not dropped. In some of the deployment
environments that we encountered, especially indoor environments, the
probability of link outage turned out to be quite large; for instance, in the case of the interiors of our department building, it turned out that even for small link lengths ($\leq 8$ meters),
about $20\%$ of the links had outage probability in excess of 5\%, and
the situation was even worse for longer links (see Figure~\ref{plot:building-pout-graph}). 

In view of this, we adopt the following design strategy: \emph{given $q_{\max}$, we first choose $h_{\max}$ so as to make $D^{(h_{\max})}_{q_{\max}} (d_{\max})$ close to 1, say 0.9999}. For example, for $q_{\max}=0.05$, and $d_{\max}=200ms$, $h_{\max}$ turns out to be 6 to ensure $D^{(h_{\max})}_{q_{\max}} (d_{\max})\geq 0.9999$. We thus ensure that when the path is not in outage,
we deliver the packets \emph{in time} with very high probability. Let us denote this choice of $h_{\max}$ as $h_{\max}^{(1)}$. Now the achievable $p_{\mathrm{del}}$ gets governed only by the $P_{\mathrm{out}}$. Then, to ensure the target delay bounded packet delivery probability, $p_{\mathrm{del}}$, we choose $h_{\max}$ such that $(1 - P_{\mathrm{out}})^{h_{\max}}\times 0.9999\geq p_{\mathrm{del}}$, i.e., 
\begin{align}
h_{\max}&=\left\lfloor\frac{\ln(p_{\mathrm{del}}/0.9999)}{\ln(1 - P_{\mathrm{out}})}\right\rfloor\nonumber\\
&\eqqcolon h_{\max}^{(2)}\nonumber 
\end{align}
Finally, the hop constraint is chosen as $\min\{h_{\max}^{(1)},h_{\max}^{(2)}\}$. For example, for $P_{\mathrm{out}}=0.05$, and $p_{\mathrm{del}}=0.77$, it turns out that $h_{\max}^{(2)}=5$, so that the final hop constraint becomes $h_{\max}=\min\{h_{\max}^{(1)},h_{\max}^{(2)}\}=\min\{6,5\}=5$. 

\subsubsection{The network design problem}
With the above mapping from $p_{\mathrm{del}}$ and $d_{\max}$ to
$h_{\max}$, our original QoS constrained network design problem can be
reformulated as the following graph design problem:

Given the network graph $G= (Q\cup P, E)$ consisting of the set of
source nodes $Q$ (including base station), the set of potential relay
locations $P$, and the set of all feasible edges $E$, \emph{extract
from this graph, a subgraph spanning $Q$, rooted at the BS, using a
minimum number of relays such that each source has at least $k$ node
disjoint paths to the sink, and the hop count from each source to
the BS on each path is $\leq h_{max}$.}  In
\cite{bhattacharya-techreport}, this is called the \emph{Rooted
Steiner Network-$k$ Connectivity-Minimum Relays-Hop Constraint}
(RSN$k$-MR-HC) problem.

For the special case of $k=1$, in \cite{bhattacharya-techreport} this
is called the \emph{Rooted Steiner Tree-Minimum Relays-Hop Constraint}
(RST-MR-HC) problem.

\subsection{Network Design Algorithms: The Basic Principle}

The details of the network design algorithms for the RST-MR-HC and the
RSN$k$-MR-HC problems are discussed in \cite{iwqos} and
\cite{bhattacharya-techreport}.

Both the algorithms basically perform a series of shortest path
computations from each source to the sink, starting with an initial
feasible solution, and adopting a certain combinatorial relay pruning
strategy to prune relay nodes from the feasible solution sequentially,
each time computing a new shortest path involving only the remaining
nodes, in an attempt to minimize relay count, while still retaining
hop count feasibility.

\subsection{Network Performance Analysis}

The final step of our design approach is to use an analytical model to
obtain the maximum packet rate that each sensor can generate so that
the QoS target is not violated for the packets generated by any
sensor.  For this we utilize a fast and accurate analytical model for
multihop CSMA/CA networks that we have reported in
\cite{winet.srivatasava-etal12performance-analysis}. Our approach
models CSMA/CA as standardized in IEEE~802.15.4, buffers at each
transmitter, and packet error rates on each wireless hop.


\section{Implementation, Practical Issues, and Testing}
\label{sec:impl-issues-testing}

\subsection{SmartConnect: System Implementation}
\begin{figure}[t]
\centering
\includegraphics[scale=0.35]{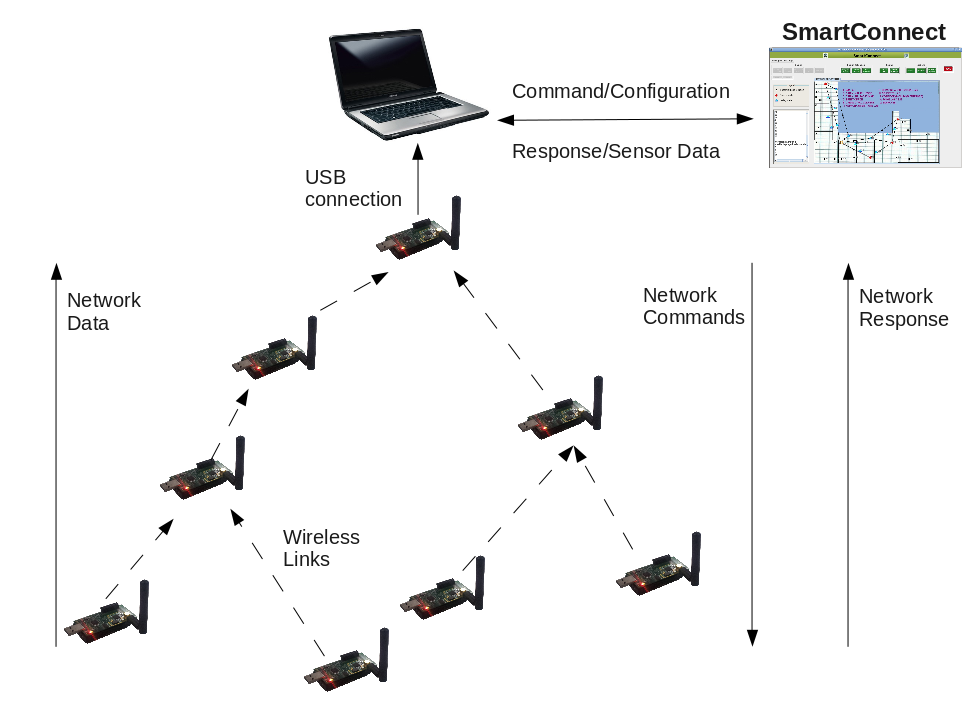}
\caption {SmartConnect architecture: Iterative deployment involves command-response interaction  between nodes on the field and the SmartConnect system via the SmartConnect-WSN gateway; message interaction between the SmartConnect graphical user interface (GUI) and SmartConnect-WSN gateway takes place over TCP/IP.}
\label{fig:SmartConnect-architecture}
\vspace{-5mm}
\end{figure}

The components of the SmartConnect network design and deployment architecture and the interactions between them are depicted in Figure~\ref{fig:SmartConnect-architecture}. The SmartConnect GUI runs the algorithms for network design and analysis at the back-end, and has a console for configuration, field interaction, and information display. The SmartConnect GUI is connected over TCP/IP to the SmartConnect-WSN gateway. The SmartConnect-WSN gateway is a Linux host connected to a base station mote over USB.

The SmartConnect GUI automates the entire design and deployment process by requiring the user to provide minimal inputs. Active participation from the user is solicited only at the time of placing relay nodes on the field. Apart from using the design algorithm to provide relay locations for
deployment, the user can intervene with their own intuitively provided
relay locations or modify those provided by the design algorithm. 
The user can also view predicted and field learnt links from each node, as well as the full network graph on which the algorithm is working in each iteration.
In our implementation the wireless nodes including the base station mote were TelosB motes with a 2.2~dBi external omnidirectional antenna for increased radio range.

\subsection{Some Practical Issues}
\label{sec:practical-issues}
\vspace{2mm}
\noindent
\textbf{Communication Before Network Set-Up}: In the design phase of
the deployment we have source nodes and relay nodes at predicted
locations on field. There is no existing reliable network connecting
the nodes on the field to the sink. At this stage, for sending
commands and receiving data from the nodes we need a protocol for
topology-free routing. The commands sent from the base station to the
nodes during link learning, and the data sent from nodes to the base
station containing link data are all sent using a form of flooding
implemented in TinyOS called dissemination
\cite{dissemination-tep118}.

When the initial design is deployed on the field, the design is based
only on modeled links. In such a deployment dissemination could fail
due to one or more `bad' links disconnecting a section of the
network. We thus cannot assure reliability of the commands being sent
out to the nodes. Since our communication range is conservative in
most cases the number of relay nodes placed on field are in excess of
what are needed and used. We depend on this aspect of our design to
provide a communication framework before the actual network comes into
place. Barring very small number of exceptions, we found that in most
test cases we could successfully reach all nodes in the network.

\noindent
\textbf{Stopping with declaration of infeasibility}: The topology
design algorithm uses either, a model based network graph (for the
initial design) or a hybrid network graph of modeled links and
field-learnt links (relay augmentation) to propose relay
locations. The link model we use is conservative so as to reduce the
number of iterations the design would require (discussed in
Section~\ref{sec:link-modeling}). So, when a model based network graph
is fed to the topology design algorithm, it could declare the design
infeasible (initial design/augmentation not possible) even
though, on field measurement of link quality, the design could be
feasible. This could happen during initial design, or when
augmentation is needed. However, in our experience, this situation did not arise over a large number of deployments. Nevertheless, to still be able to attempt a design, even when the design algorithm declares infeasibility, SmartConnect allows the user to intuitively place relays at any potential relay location,
and proceed with the design process. On evaluation of the field-learnt
network graph the design may or may not be successful. The user is
allowed to continue augmenting the network and evaluating until relays
are placed at all potential relay locations. 

\subsection{Testing}
After network design and validation, a test phase is conducted to verify
whether the delivery probability and delay promised by the design are being met
by the network. In this phase the network is essentially in operation, but the
data being relayed by the network are pseudo-sensor data. In the specific case that we
are addressing, the data from each sensor is collected infrequently. The time
duration required to receive from each sensor source depends on the number of
hops it is from the base station and is in the order of only a few milliseconds. This
allows us to collect data from each source in a Time Division Multiplexed (TDM)
manner. A time frame is created with one slot given to each source. Each
source sends in its time slot, thus maintaining light traffic in the network. This
requires us to have a time sync protocol in place, so the time frames of all
nodes are synchronized. The time sync protocol used in our implementation is
FTSP (flooding time synchronization protocol) provided by TinyOS.


\section{Experiences with Experimental Deployments inside our
  Department Building}
\label{sec:deployments-results-experiences}

SmartConnect has been used to make test deployments in three very
different environments: inside our department building, on the lawns
of our building, and at a large power distribution yard (recall
Figure~\ref{fig:nelamangala-location}). We found that by far the most
challenging environment was the building. Hence, we report here our
experience with a test deployment in our department.

\subsection{Indoor Deployment 1}

We use this test deployment, to test the basic working of the
iterative design and the approximation algorithms. GUI snapshots are
shown at each design phase.

This deployment environment is a high ceilinged building, constructed
from stone blocks (granite), and built during the period 1946 to 1951. It
has thick stone walls (0.66~m thick outer walls, and 0.66~m thick walls
between rooms), 5m high ceilings, heavy wooden doors, cubicle
separators, tables and other office equipment. As described in
Section~\ref{sec:link-modeling} a communication range model was
obtained for the environment before the deployment. Owing to heavy
attenuation by walls and doors with very few corridors, link outage in
the environment was large, resulting in an $R_{\mathsf{max}}$ value of
only 8~m. 

The deployment parameters of this test deployment are as
follows. Field area of 1650 $m^{2}$, 24 potential locations, 10 sensor
sources, 200~ms delay constraint, communication range of 8~m and a
path redundancy of 1. The target, delay bounded, packet
delivery ratio $p_{\mathsf{del}}$ for this network deployment was 73\% which would allow a network of at most 6 hops (recall from Section~\ref{sec:network-design-approach}, the method of choosing $h_{\max}$).

\begin{figure}[p]
\begin{center}
  \includegraphics[scale =
  0.27]{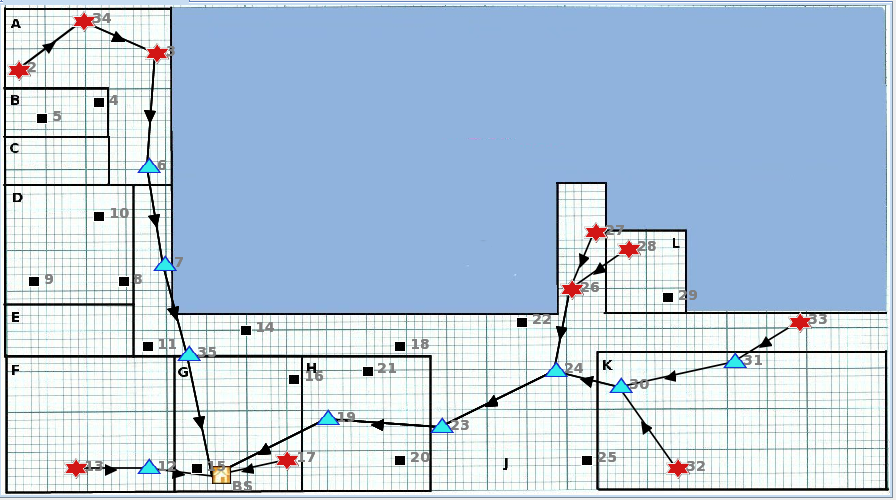}
\end{center}
\caption{Indoor deployment 1: Initial design on the model based
  network graph. 10 sources; the initial model-based design suggests 9
  relays; the paths in the initial design are shown.}
\label{fig:building_design_phase}


\begin{center}
\includegraphics[scale = 0.27]{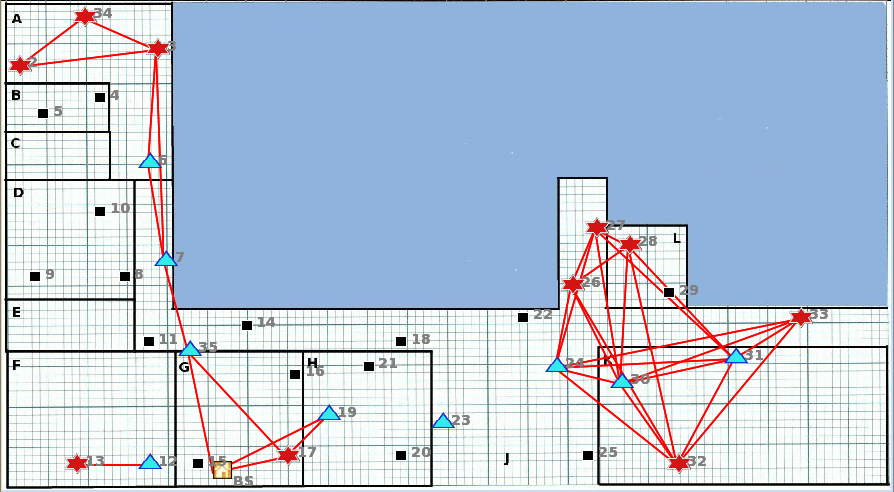}
\end{center}
\caption{Links learnt after deploying the relays suggested by the initial design. All good links learnt are shown.}
\label{fig:building_iteration1_links}


\begin{center}
\includegraphics[scale = 0.27]{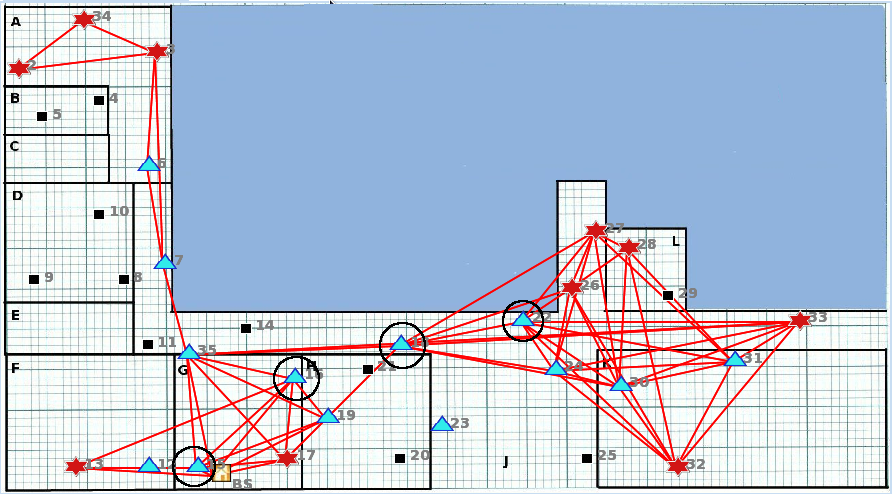}
\end{center}
\caption{Augmentation step suggests the placement of Relays 15, 16, 18, and 22. The additional good links learnt now yield a connected network.}
\label{fig:building_iteration2_links}


\begin{center}
\includegraphics[scale = 0.27]{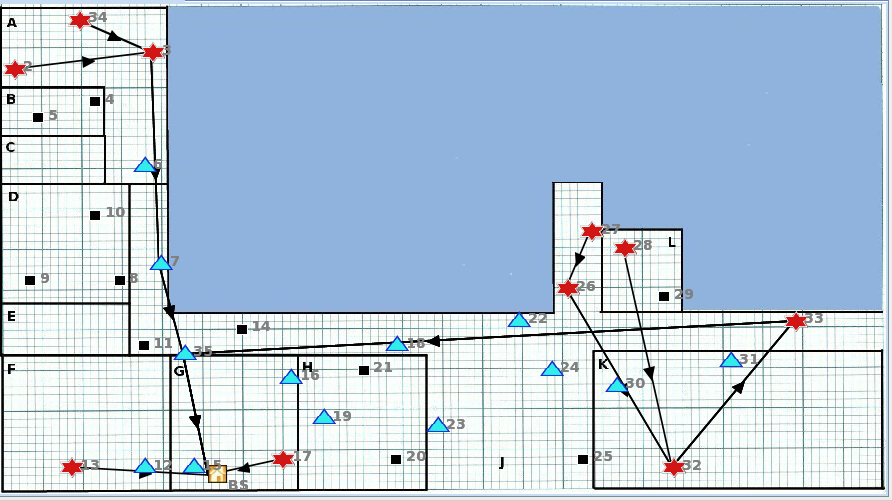}
\end{center}
\caption{Final network design based only on the good links learnt in the field. Just two of the relays deployed in the field (namely, 7, and 35) end up being needed. $\lambda_{max}$ of the network is 0.103 pkts/s.}
\label{fig:building1_final_paths}
\vspace{-8mm}
\end{figure}

\begin{table}[t]
\begin{center}
\begin{tabular}{|c|c|c|}
\hline
Sensor	 &   $Measured P_{del}$   &   Predicted $P_{del}$ \\
source ID &   &  \\
\hline
2		&	   0.9751    	&  0.9119\\  
3    		&	   0.9857	&  0.9448\\	
34    		&	   0.9791	&  0.9163\\
13    		&	   0.9997       &  0.9518\\
17    		&	   1.0000	&  0.9880\\
26    		&	   0.9191	&  0.9122\\
27    		&	   0.8548	&  0.9120\\
28    		&	   0.9389       &  0.9155\\
32    		&	   0.9581    	&  0.9259\\
33		&	   0.9911	&  0.9345\\
\hline
\end{tabular}
\end{center}
\vspace{-1mm}
\caption{Indoor Deployment 1: Delay bounded packet delivery rate in the final design shown in Figure~\ref{fig:building1_final_paths}.}
\label{table:results_deployment1}
\vspace{-9mm}
\end{table}

Figure~\ref{fig:building_design_phase} shows a GUI snapshot of the
initial design on the model based network graph. The sensor sources
are indicated by red stars, the black squares are potential relay
locations, the yellow house is the sink, and the blue triangles are
relay nodes. The design algorithm suggested relay locations are
indicated by blue triangles and the paths shown are QoS abiding paths
on the model based network graph. We see that the initial design
suggested nine relays (numbered 6, 7, 35, 12, 19, 23, 24, 30, and
31). The GUI snapshot also shows the paths that each source should
use.

After placing relays at locations suggested by the topology, link
learning was done with the nodes on field. The field-learnt links
between these nodes are shown in
Figure~\ref{fig:building_iteration1_links}. Each red line on the graph
is a bidirectional \emph{`good'} link. A link is said to be
bidirectional if link outage constraints are met when measured in
either direction. We see from the figure that the learnt-links network
graph was not even fully connected. Evaluation of this field-learnt
network graph failed, and a second iteration was required with the
design algorithm suggesting augmentation of relays (four relays,
numbered 15, 16, 18, and
22). Figure~\ref{fig:building_iteration2_links} shows the field-learnt
links after augmentation and the second iteration of link learning
(the augmented relays are highlighted with a circle). This graph on
evaluation was found to meet QoS. The final design is shown in
Figure~\ref{fig:building1_final_paths}. Some observations from this
design are that; of all the relays suggested, only two relays
(numbered 7 and 35) were used by the design (relays 6, 12, 15, 16, 19,
18, 23, 22, 24, 30, and 31 were removed after the design); some
sources are also acting as relays in the design; the link between
nodes 33 and 35, even though very long, was learnt to be `good' on
field since it had a clear line of sight path. An important observation is that a link that was not there in the first iteration appeared later, indicating link instability. This is addressed later in Section~\ref{sec:robustness}.

Once the design was complete, we ran the analytical model described in
\cite{winet.srivatasava-etal12performance-analysis}, and found that to
meet the QoS requirement, the maximum packet generation rate from any
sensor, for a Poisson packet generation process, i.e.,
$\lambda_{\max}$, is 0.103 pkts/s; i.e., about 1 packet every 10
seconds from each sensor, which is quite adequate for applications
such as condition monitoring.

Finally, field testing was performed by sending pseudo-sensor data
packets over the network. Results are provided in
Table~\ref{table:results_deployment1}. Predicted $p_{\mathsf{del}}$ 
for each source node in the table was found by using field-learnt 
link outage values in the same $p_{\mathsf{del}}$ inequality. 

We see that Node 27 has a
delivery probability 6\% lower than the predicted value. Due to the
dynamic nature of this environment, after deployment, some links turned out to be worse
than measured. This could be due to the movement of
people around some motes or opening/closing of doors. In this example the predicted and 
the measured delivery probabilities are still better than the target $p_{\mathrm{del}}$, 
which we recall was 73\%. However, this need not continue to hold over time, 
as long term variations in the statistics of the links affects delivery rates. 
This motivates the need for a repair
phase (refer to the last block of the flow diagram in
Figure~\ref{fig:iterative-design-process}) that is triggered by a drop
in delivery rate to handle long term variations and make the network
robust. The procedure for doing this along with some experimental
results are discussed in Section~\ref{sec:robustness}.

\subsection{Indoor Deployment 2}

\begin{figure}[t]
\begin{center}
\includegraphics[scale = 0.27]{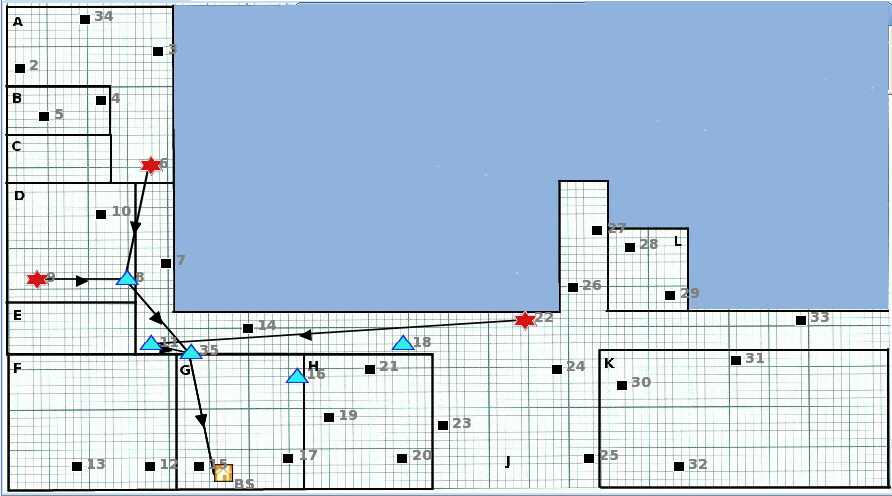}
\end{center}
\caption{Indoor deployment 2: Three sensors; five relays proposed by
  the initial design (namely, 8, 11, 35, 16, 18); finally just three
  relays (8, 11, and 35) are used. $\lambda_{max}$ of the network is 0.118 pkts/s.}
\label{fig:building_deployment2}
\vspace{-2mm}
\end{figure}

\begin{table}[t]
\begin{center}
\begin{tabular}{|c|c|c|}
\hline
Sensor    &    Measured  $p_{\mathsf{del}}$    &  	Predicted $p_{\mathsf{del}}$\\
source ID & & \\
\hline
6 & 0.8971 & 0.8830\\
9 & 0.8961 & 0.8930\\
22 & 0.9075 & 0.9300\\ 
\hline
\end{tabular}
\end{center}
\vspace{1mm}
\caption{Indoor deployment 2: Delay bounded packet delivery rate in the final design shown in Figure~\ref{fig:building_deployment2}.}
\label{table:building-deployment-2}
\vspace{-5mm}
\end{table}

The results of another smaller deployment made in our department
building is presented here. The design parameters for this deployment
were the same as in the previous example, except that 3 sensor sources
were deployed here and the target $p_{\mathsf{del}}$ for this deployment was 77\% allowing at most a 5 hop network (see Section~\ref{sec:network-design-approach} for details of the procedure to choose $h_{\max}$).

Figure~\ref{fig:building_deployment2} shows a snapshot of the final
network design. This network required only one iteration of design and
evaluation. Of the five relays suggested by the initial design, three (namely, 8, 11, and 35) were used and the other two (namely,
16, and 18) were removed. On completion of the design, we analyzed the network performance for positive traffic arrival rates as in the previous example, and found that to meet the QoS requirement, the maximum packet generation rate, i.e.,
$\lambda_{\max}$, from any sensor, for a Poisson packet generation process, is 0.118 pkts/s. The results of field testing on the network are shown in Table~\ref{table:building-deployment-2}. Notice  that for Node 22, even though the measured $p_{\mathsf{del}}$ was worse than the predicted $p_{\mathsf{del}}$, both were still much better than  the target, i.e., 77\%. 
\subsection{Large Deployment: Outdoor and Indoor}
\begin{figure}[p]
\centering
\begin{center}
\includegraphics[scale=0.23]{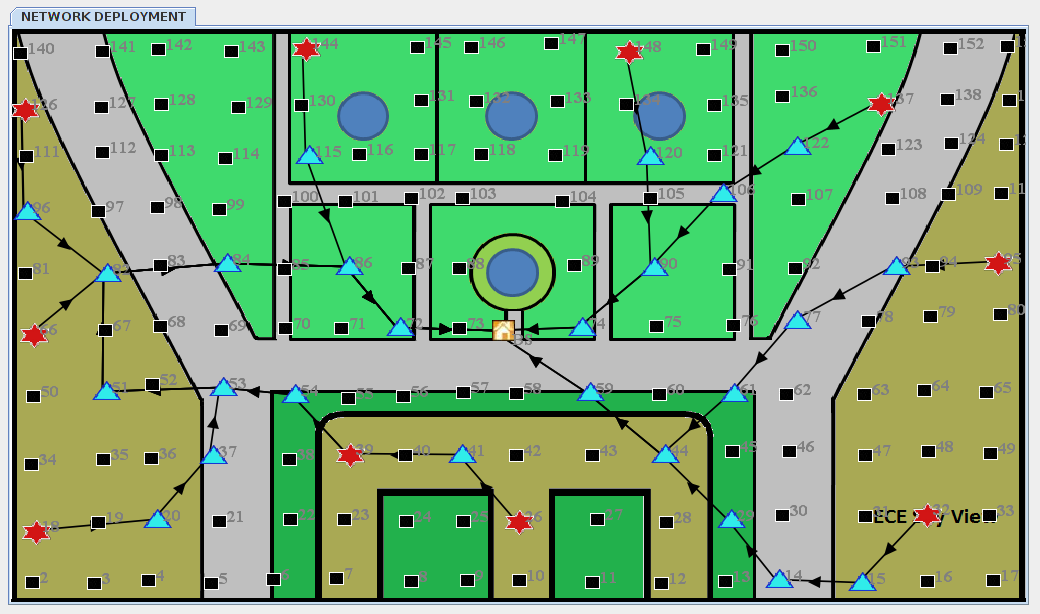}
\caption{Large deployment. Initial design on the model based network graph. 10 sources; the initial model-based design suggests $25$ relays; the paths in the initial design are shown.}
\label{fig:outdoor_deploy_routes}
\end{center}

\begin{center}
\includegraphics[scale=0.23]{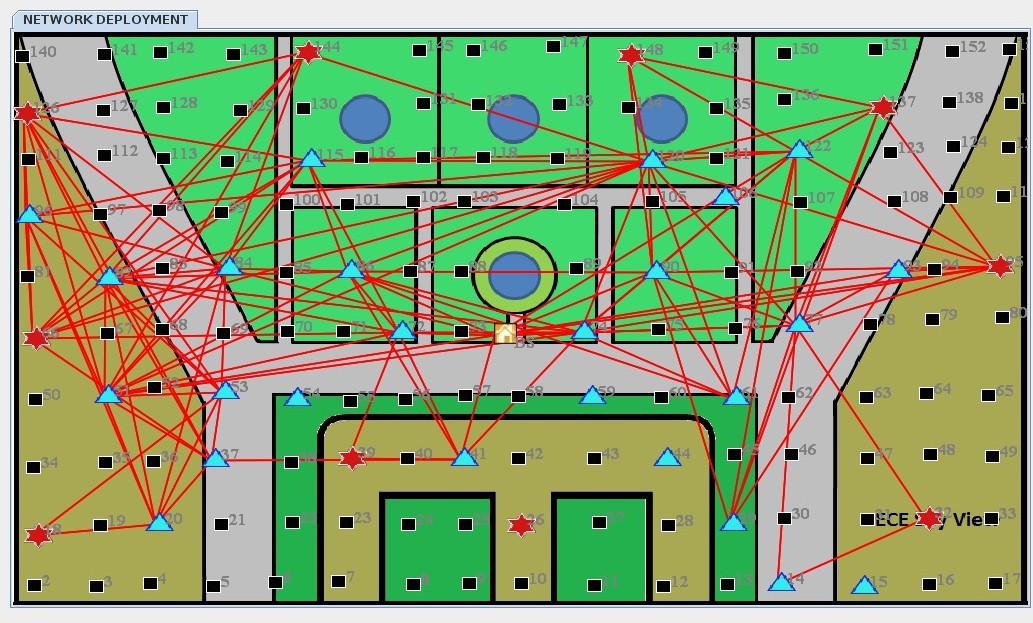}
\caption{Links learnt after deploying relays at the suggested locations. All good links are shown}
\label{fig:outdoor_deploy_it1}

\end{center}

\begin{center}
\includegraphics[scale=0.23]{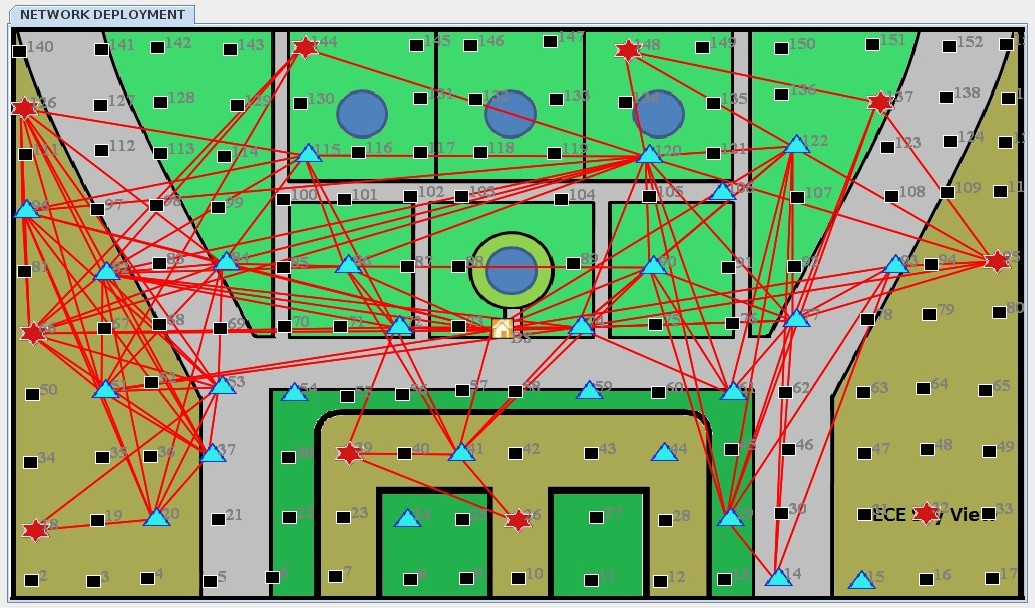}
\caption{Links learnt after Relay 24 is deployed as suggested for augmentation. However, source 32 is disconnected.}
\label{fig:outdoor_deploy_it2}

\end{center}

\begin{center}
\includegraphics[scale=0.23]{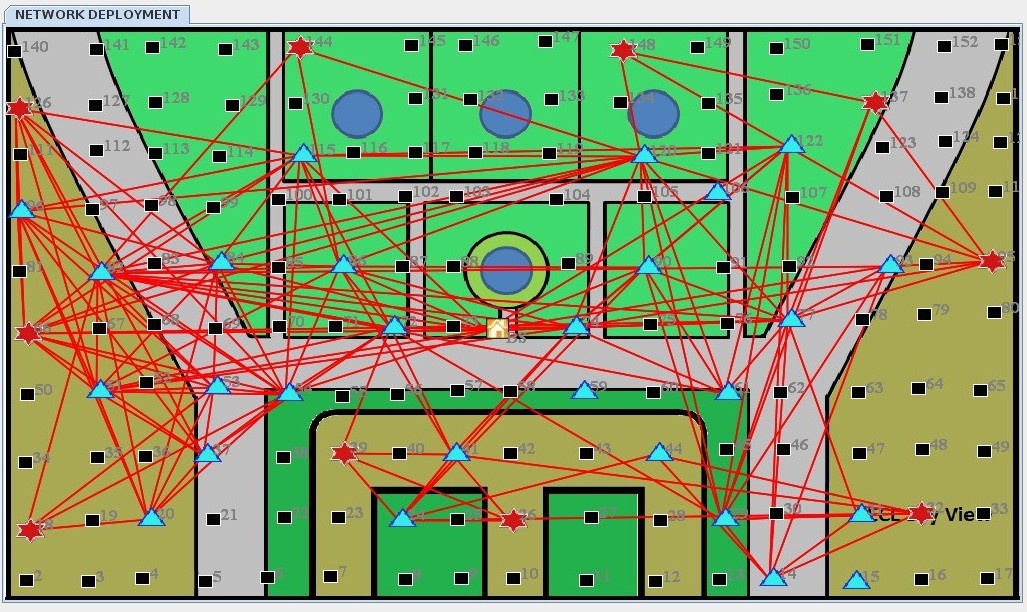}
\caption{Links learnt after iteration 3 with  Relay 31 deployed as suggested for augmentation.
The links yeild a connected network.} 
\label{fig:outdoor_deploy_it3}
\end{center}
\end{figure}

\begin{figure}
\begin{center}
\includegraphics[scale=0.23]{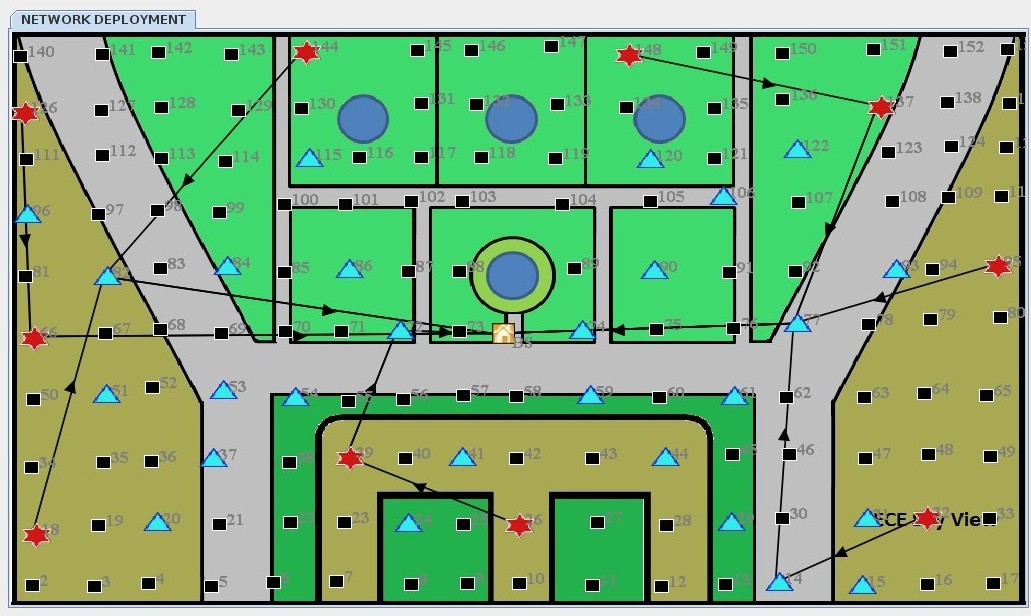}
\caption{Final network design based only on the good links learnt on the field. Just four of the relays deployed on the field are used.}
\label{fig:outdoor_final_design}
\end{center}
\end{figure}

We further test the tool with larger field dimensions. This deployment is performed on the ECE department premises of IISc which constitutes the building described above (indoor deployment) along with lawns with thick vegetation and parking lot on either side of the building with vehicles parked(outdoor deployment). The deployment parameters are as follows: Field area of $8750m^2$, $153$ potential locations, $10$ sensor sources, $250 ms$ delay constraint, communication range of $15m$(the $R_max$ value for outdoor environment) and a path redundancy of $1$. 

Figure \ref{fig:outdoor_deploy_routes} shows the GUI snapshot of intial design on the model based graph. The colour conventions for the sources, relays and the base station are as described for the indoor deployment in Section \ref{sec:deployments-results-experiences}. The initial design suggested $25$ relays. The paths that each source should use are also indicated in the figure.

As described and discussed in the indoor deployment, link-learning is done with the nodes on field after placing the relays at the suggested locations. Figure \ref{fig:outdoor_deploy_it1} shows the field learnt links for the given deployment. It can be observed that one source no. $26$ (Bottom-center in the figure) has no `good' links from any of the other nodes. Hence evaluation of the field-learnt graph failed and augmentation was suggested. Figure \ref{fig:outdoor_deploy_it2} shows the field-graph for the second iteration. However, now source no $32$ (Bottom-right in the figure) is disconnected (indicating link instability) and another augmentation was suggested. After the third iteration, all the source nodes are connected to the base station while meeting the required QoS. The calculated paths after the final successful design are shown in Figure \ref{fig:outdoor_final_design}.
\section{Robust Network Design}
\label{sec:robustness}

\begin{figure}[t]
\centering
\includegraphics[scale=0.34]{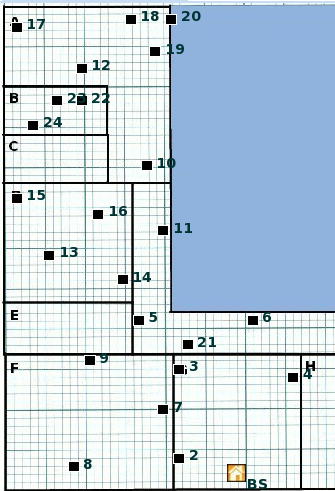}
\caption{Set of 24 locations where nodes were placed and link learning
  performed periodically to evaluate network robustness}
\label{fig:robustness-locations}
\vspace{-3mm}
\end{figure}

\begin{table}[t]
\scriptsize
\begin{center}
\begin{tabular}{|p{4mm}|c|c|p{17mm}|p{12mm}|p{10mm}|}
\hline
Expt.    &    Source IDs     &   Initial     &    Augmentation required &  Final relay &  no. of topology \\
  no.      &                   &     relay set &    at cycles            &    set       & redesigns\\
\hline
           &                &               &                         &              &\\

1 & 23, 11  & 10, 3     & 6, 7 & 10, 3, 22 & 5\\
  & 9, 12   &           & 17   & 2, 5    & \\
\hline
2 & 18, 19 & 10    & 17 & 10, 11 & 4\\
  & 6, 4   &       &    &  5     & \\
\hline
3 & 20, 22 &  no relays &  none & none & 0\\
  & 5, 21  &  were used &       &   & \\
\hline
4 & 18, 22 & 11, 3  &  none & 11, 3 & 1\\
  & 16, 6  &        &       &   & \\
\hline
5 & 15, 10 & 3  & none & 3 & 0\\
  & 11, 4  &    &      &   & \\
\hline
6 & 19, 20&  11 & none & 11 & 0\\
  & 21, 3 &  &  & & \\
\hline
7 & 12, 15 &  14, 5& 12, 32 & 14, 5 & 8\\
  & 21, 3  &       &        & 10, 11 & \\
\hline
8 & 23, 24 &  14, 5 & 16, 17  & 14, 5, 3, 10 & 3\\
  & 15, 16 &  3     &         &  12, 11, 6   & \\
\hline
9 & 18, 24 &  10, 3 & 17, 20 & 10, 3 & 2\\
  & 11, 5  &        &        & 6, 12 & \\
\hline
10 & 10, 6  &  no relays  &13, 17 & 3, 11 & 3\\
   & 12, 19 &   were used &       & 5     & \\
\hline
\end{tabular}
\end{center}
\caption{Results from the experiment to study temporal robustness
  of the network design, where at each design stage only one path is ensured (i.e., $k=1$); see text for details.}
\label{table:network-robustness}
\normalsize
\vspace{-7mm}
\end{table}

\begin{table}[t]
\scriptsize
\begin{center}
\begin{tabular}{|p{4mm}|c|c|p{17mm}|p{12mm}|p{10mm}|}
\hline
Expt.    &    Source IDs     &   Initial     &    Augmentation required &  Final relay &  no. of topology \\
  no.      &                   &     relay set &    at cycles            &    set       & redesigns\\
\hline
           &                &               &                         &              &\\

1 & 23, 11  & 10, 3, 22       & no augmentat- & 10, 3, 22 &1 \\
  & 9, 12   &  14, 1          & -ion required        & 14, 1     & \\
\hline
2 & 18, 19 & 10, 22, 14    & no augmentat- & 10, 22, 14 & 2\\
  & 6, 4   & 5, 3          & -ion required        & 5, 3       &   \\
\hline
3 & 20, 22 & 10, 6  & no augmentat-   & 10, 6 & 0\\
  & 5, 21  &        & -ion required          &       & \\
\hline
4 & 18, 22 & 10, 11   & 13  & 10, 11  & 1\\
  & 16, 6  &  3       &     & 3, 4    &\\
\hline
5 & 15, 10 & 3, 16  & no augmentat-  &  3, 16 & 0\\
  & 11, 4  & 6      & -ion required         & 6      & \\
\hline
6 & 19, 20 & 5, 11  & no augmentat- & 5, 11 &0 \\
  & 21, 3  &        & -ion required        &       & \\
\hline
7 & 12, 15 & 14, 5  & no augmentat- & 14, 5  & 0\\
  & 21, 3  & 10, 11 & -ion required        & 10, 11 & \\
\hline
8 & 23, 24 & 5, 3, 12   & no augmentat- & 5, 3, 12 & 0\\
  & 15, 16 & 10, 6      & -ion required        & 10, 6    & \\
\hline
9 & 18, 24 & 10, 3  & no augmentat-  & 10, 3  & 0\\
  & 11, 5  & 15, 21 & -ion required         & 15, 21 & \\
\hline
10 & 10, 6    & 22, 14 & 13, 17& 22, 14, 5 & 2\\
   & 12, 19   & 5, 3   &       & 3. 21, 11 & \\
\hline
\end{tabular}
\end{center}
\caption{Results from the experiment to study temporal robustness
  of the network design, where at each design stage two node disjoint paths are ensured from each source (i.e., k=2); see text for details.}
\label{table:network-robustness-k2}
\normalsize
\vspace{-9mm}
\end{table}

The SmartConnect design approach accounts for short term link
variations (over channel coherence times) via the wireless link
model. A sensor network would, however, be expected to operate over a
long period of time; at least months, and even years. Over such long
periods, due to changes in the propagation environment (e.g., for
in-building networks, the changes could be new furniture, partitions, etc.), the quality of the links in the original design could significantly vary causing a
decrease in packet delivery rates. The on-field iterative approach we
have adopted, permits us to easily address the problem of network
robustness under long term variations in the links; see the design
flow chart in Figure~\ref{fig:iterative-design-process}.

After the initial design has been done, network operation starts, and
the network monitors the packet delivery rates. \emph{Repair} of the network
is initiated when the delivery rate of data reduces below what it was
designed for. At this point the links between the nodes in the field
are learnt again, and, if possible, the topology is redesigned with no
addition of relay nodes. If this is not possible then, exactly as in
the initial design process, the modeled links are included and a
proposal is made for augmentation with additional relays. At each
augmentation stage, \emph{all the heretofore deployed relays are retained}. The reason for doing this is to provide robustness against future channel variations so that such variations can be handled just by topology redesign over the existing nodes, without needing further relay augmentation.

With this approach in mind, it is interesting to ask the following questions:
\begin{enumerate}
\item How often is topology redesign required?
\item Does it indeed happen that after some time no further relay
  augmentation is required?
\item Does it help to design the network with path redundancy?
\end{enumerate}

In order to study these questions we carried out an experiment in
which a set of 24 locations were identified inside a section of our
department building (Figure~\ref{fig:robustness-locations}). Unlike
the deployment experiments reported in
Section~\ref{sec:deployments-results-experiences}, we confined these
locations to the several offices and labs occupied by Network Labs (ECE Department, IISc) (about 600 square meters) in order to be
able to leave the relays undisturbed for several days. With the nodes
at these 24 locations, link learning was performed periodically with a
gap of 4 hours, 20 times over a span of 3 days, and another 20 times
over a span of another 3 days, a week later; a total of 40 evaluation
and possible redesign cycles.

With the link quality data collected, we could then study (offline) the effect of link variations and the presense of alternate paths in networks designed over
these 24 nodes. To study the approach where the network is designed
with only one path from each source to the basestation (i.e., k=1), we considered 10 sets of 4 nodes as sources, and designed networks (for a target delay of 200 ms) connecting these sources to a base station, taking the remaining 20 locations as potential relay placement points, using the proposed iterative design approach in the first evaluation cycle. For each of these 10 network design problems, by using the collected link quality measurements we could track the evolution of the delivery probabilities over the 40 evaluation cycles, and (virtually) carry out topology redesign and relay augmentation. A redesign was triggered when the delivery rate from any source (as estimated from the measured qualities of the links being used in each design) dropped to below 73\% (which is the least delivery rate expected for a 6 hop network with outage $\leq$ 5\% along each link (discussed in Section~\ref{sec:qos-to-hop-constraint})). The performance deterioration would then be attempted to be resolved by topology redesign or by relay augmentation.

We report our results of such an experiment with the network being redesigned at each repair stage so that there is one path connecting each source to the basestation (i.e., $k=1$), in Table~\ref{table:network-robustness}. The
second column of this table shows the 10 different sets of nodes that
were sources in the 10 experiments (the numbers relate to
Figure~\ref{fig:robustness-locations}). The third column shows the set
of relays used in the initial design. In the fourth column we show the
indices of the 40 evaluation cycles at which relay augmentation was
needed. The next column shows the final set of relays, and the final
column shows the total number of times redesign (with or without relay
augmentation) was done over the 40 evaluation cycles.

From Table~\ref{table:network-robustness}, we see that a maximum of 8
topology redesigns were required over the 40 cycles. So, in the worst case, a topology redesign was required around once in a day. In all cases, except experiment 7 (which, in fact, also required the most number of repairs, 8), no augmentation was required after at most the 20th evaluation cycle, even when the network was evaluated after a full week for another 3 days (in fact, cases 3, 4, 5 and 6 never required a relay
augmentation). Hence, it appears that we eventually
converge to a deployment where no further relay augmentation will be
needed, and topology redesigns over the existing nodes alone will
suffice to take care of long term variations. Note that, in this
approach, we are essentially over-deploying relays to create redundant
links for robustness of the deployed network. But it is also important
to notice that in most cases the difference in the size of the initial
relay set and final relay set is not large (the worst case being
Experiment 8 where four relays had to be added to the original three, out of a total of 20 potential relay locations)
indicating that the number of extra relays required to take care of
time varying link qualities is quite small.

Having explored this, we then began to answer the third question posed earlier in this section. At the
very beginning, if we design the network to have path redundancy (node disjoint
paths), would we reduce the number of redesigns required with time? 

The experiments were carried out in the same way as described above, except that in the first design cycle we design 2 node disjoint paths
from each source to the basestation such that the QoS is met along both the paths
for each source (i.e., in the notation of this paper, $k=2$). This design is now evaluated along the 40 cycles, and
redesign is triggered only if the delivery probabilities along both the paths of any source violate the target delivery probability of 73\%.  As in the $k=1$ experiments, we considered 10 source sets of 4 sources each. The results are presented in Table~\ref{table:network-robustness-k2}. We see that no augmentation was required to the initial relay set in 8 out of 10 cases as compared to 4 out of 10 cases without path redundancy. The maximum number of topology redesigns required was 2 whereas it was 8 in cases without path redundancy. Also, no redesign
was required beyond the 17th evaluation cycle. The maximum number of relays
added in cases where augmentation was required was also just 2 relays. While designing with path redundancy, overdeployment is done in the very first design step, causing the network to converge faster than with no path redundancy. In
comparison with the approach in which the network is initially designed with only one path from each source to the sink, we
conclude that adding path redundancy at the very beginning significantly improves the robustness of the network to long term channel variations.

\section{Experiments with a Dynamic Routing Protocol: RPL on SmartConnect}
\label{sec:rpl}
Having designed a QoS aware, robust network using SmartConnect, it is interesting to explore how the network performance changes when operated in conjunction with some routing protocol that selects routes \emph{dynamically} to optimize some predefined objective (such as end-to-end delivery probability) instead of using the static routes suggested by SmartConnect. In particular, we are interested in studying the behavior of the networks obtained using SmartConnect when used in conjunction with RPL \cite{phinney}, which is an industry standard routing protocol for wireless networks with time-varying and lossy links. RPL can be programmed to use any objective function, and link quality metrics. For our purposes, we define the objective function to be to maximize the end-to-end delivery probability, and the link quality metric to be the packet error rate on a link. When the nodes suggested by the SmartConnect design are deployed on field, and RPL is run on them, RPL  monitors the links, and accordingly, keeps updating the link quality estimates of the links, and dynamically selects the best routes (basically using Bellman-Ford algorithm with the most recent link quality estimates) in accordance with the predefined objective function. A more detailed description of the link quality estimation, and path update mechanism is as follows \cite{rpl-rfc,dawans}: each node continuously maintains a list of potential parents, one of which is chosen as the \emph{preferred parent} based on their current path qualities (``ranks'' in RPL terminology) to the sink. Routing from a node is performed through this preferred parent. The link qualities of a node to all of its potential parents (including the preferred parent) are initialized to some default value. In addition to the data traffic from the source nodes, each node sends periodic control packets (DAO messages) to its preferred parent. The link qualities are updated by measuring success rates on the links \emph{currently being used} by data or control traffic. Thus, the most recent updates are available for only those links that are part of the currently active RPL tree topology, i.e., only the links between nodes and their preferred parents.\footnote{As pointed out by Dawans et al. \cite{dawans}, ``this approach is conservative in the sense that it only evaluates the links that are currently being used. It efficiently detects link failures towards the current preferred parent, but doesn't investigate any alternative otherwise. As a result, CTP and RPL often stick to a routing topology that may become sub-optimal with time.''} An update in link quality triggers an update in the rank of the corresponding transmitting node. Whenever a node's rank is updated, it broadcasts a control packet announcing its new rank; the nodes receiving this broadcast refresh their potential parent list, and recompute their preferred parent, potentially causing an update in their ranks (and hence, possibly, their paths to the sink) as well. 

Intuitively, use of such a dynamic routing protocol with the SmartConnect based network can provide the following advantages:

\begin{enumerate}
\item Recall that SmartConnect suggests routes that are just good enough to meet the predefined QoS target. On the other hand, RPL aims at optimizing the QoS objective, not just meeting the target. Hence, on the same collection of nodes (those obtained from SmartConnect design), RPL may be able to find better routes (in terms of QoS) than those suggested by SmartConnect.
\item In SmartConnect, whenever the QoS drops below the target QoS, we have to trigger repair which results in a topology redesign (rerouting without relay augmentation), or relay augmentation. Since RPL selects routes dynamically, it may take care of topology redesigns automatically, and thus, eliminate the need for invoking the SmartConnect repair phase until such time as relay augmentation becomes necessary (i.e., no QoS satisfying routes exist in the current on-field network graph). 
\end{enumerate} 

\subsection{Experiments with $k=2$}
\label{subsec:rpl-k2}
To verify the above intuitions, we performed the following experiment.

\gap
\noindent
\textbf{The experiment:}
\begin{enumerate}
\item Using SmartConnect, we designed a network for 4 sources and $k=2$. The design parameters were the same as in the robustness experiments described in Section~\ref{sec:robustness}. The resulting topology is shown in Figure~\ref{fig:rpl-smartconnect-design}.
\begin{figure}[t]
\centering
\includegraphics[scale=0.25]{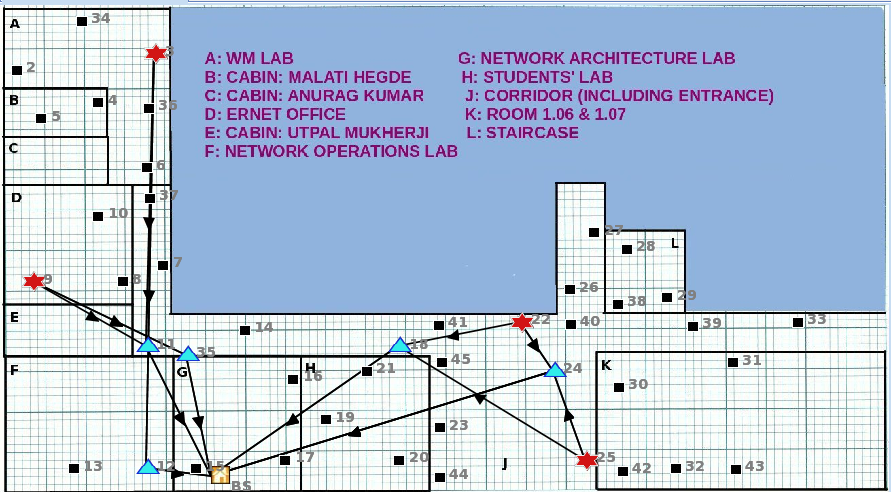}
\caption{Design obtained using SmartConnect; RPL was run after deploying relays at the suggested locations}
\label{fig:rpl-smartconnect-design}
\vspace{-3mm}
\end{figure}
\item At each of the locations (including the source locations) suggested by the SmartConnect design, two motes were deployed, one programmed for static routing (using the routes suggested by SmartConnect), and the other programmed for running RPL with PER as link quality metric, and (maximizing) end-to-end delivery probability as the objective function; the two motes at each location were operated at channels that were 30 MHz apart. 
\item To comply with the lone-packet model, each source generated traffic at a rate of 1 packet every 15 seconds. Both the RPL network, and the static routing network were operated in parallel for 5 days. 
\begin{itemize}
\item In RPL, the PER on a link was computed over a window of 20 packets, and the the link quality estimate was updated as:
\begin{align*}
\text{current estimate} &= (1-\alpha)\times \text{current PER} \nonumber\\
&+ \alpha\times \text{previous estimate}\nonumber
\end{align*}
where $\alpha = 0.5$. The initial estimates for all the links were set to 1.  
\item In SmartConnect, tracerouting was performed every 150 seconds to select one of the two static routes for each source. 
\end{itemize}
\item The end-to-end delivery probability of each source in each network was continuously monitored over windows of 100 packets. 
\end{enumerate}

\begin{figure}[t]
\centering
\includegraphics[scale=0.35]{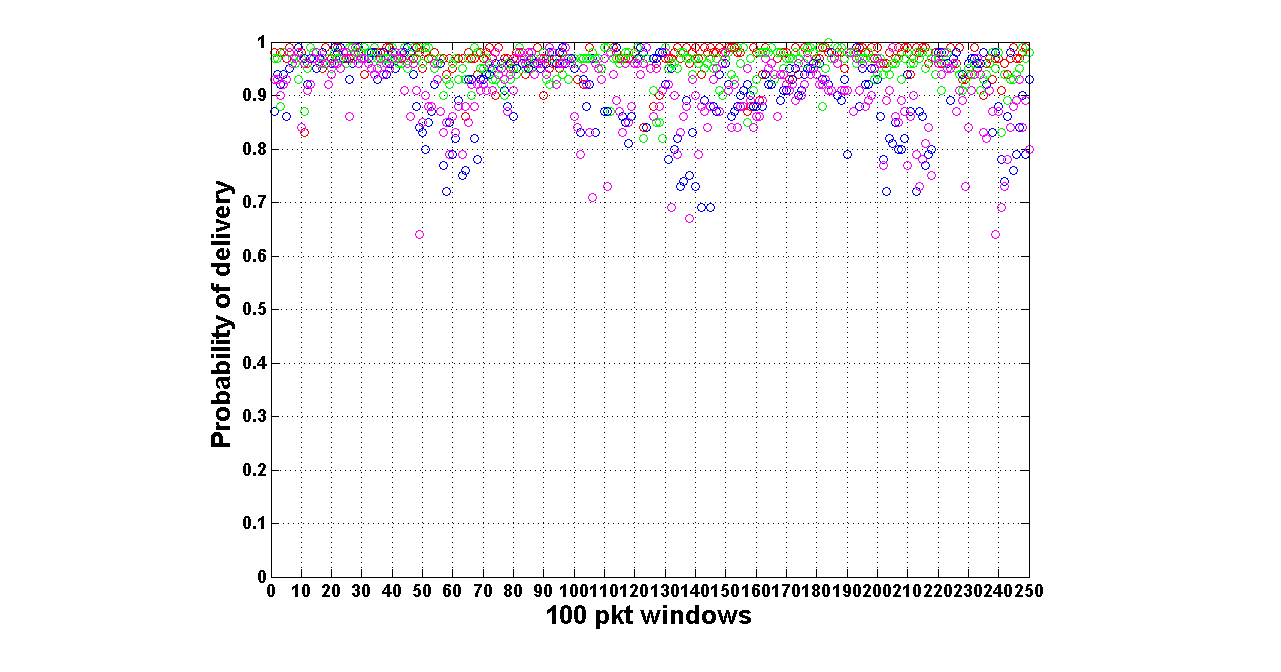}
\vspace{1mm}
\includegraphics[scale=0.35]{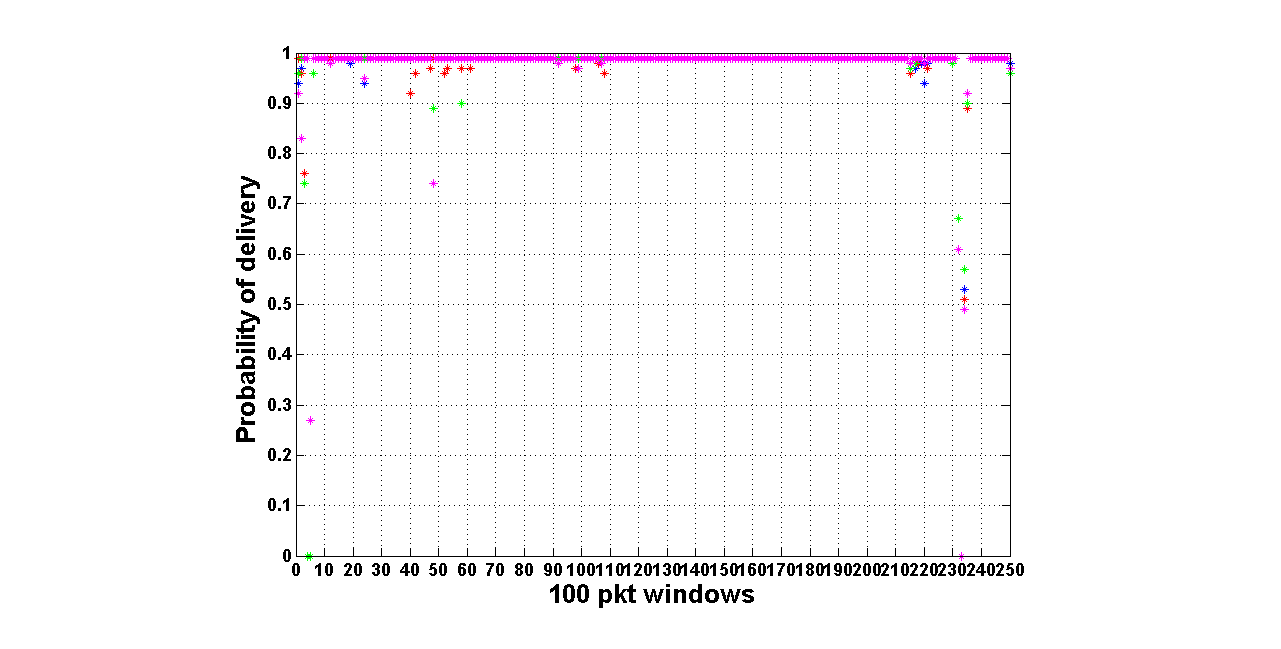}
\caption{Results of using RPL in conjunction with SmartConnect. Top panel: performance of static routing; Bottom panel: performance of RPL}
\label{fig:rpl-smartconnect}
\vspace{-3mm}
\end{figure}

The results are summarized in Figure~\ref{fig:rpl-smartconnect}.

\gap
\noindent
\textbf{Observations and Discussion:}

\begin{enumerate}
\item From Figure~\ref{fig:rpl-smartconnect}, we observe that the end-to-end delivery probability achieved by RPL over the entire period of 5 days was better than that achieved by the static routing of SmartConnect.\footnote{Note, however, that for the most part, SmartConnect still met the target $p_{\mathrm{del}}$ of 73\%.} This is expected due to the following reasons.

\begin{enumerate}
\item Since SmartConnect uses static routes, a drop in any link quality in any of these routes adversely affects the SmartConnect QoS, and the QoS cannot recover until the repair phase is triggered, or the affected links recover, whereas RPL being a dynamic routing protocol, can better adapt to the time-varying link qualities, and can automatically switch to an alternate path with better QoS without having to wait for the affected links to recover. 
\item Note that the objective function of RPL is only packet delivery probability (and not end-to-end delay), unlike SmartConnect, where the design was done to ensure a \emph{delay-bounded} packet delivery probability, and hence the design was somewhat more conservative. Note that paths that achieve the minimum $p_{\mathrm{del}}$ need not meet the delay requirement. This delay-oblivious objective function of RPL could be another possible reason why RPL does better than SmartConnect static routes in terms of $p_{\mathrm{del}}$ even during the initial windows.
\end{enumerate}
\item With RPL, the end-to-end delivery probability of any source seldom dropped below 90\%, and even when it did, it recovered within at most 1 window, i.e., 100 packets (approximately 25 minutes). Note that this recovery is \emph{non-disruptive}, i.e., network operations may continue during the recovery process unlike SmartConnect repair phase, which, if invoked, would shut down network operations until the repair is complete. Further note that the repair phase involves at least one round of link learning, which, for a 9-node network as in our experiment, would take about 30 minutes, and hence is, at best, comparable to the recovery time of RPL. Thus, RPL can automate topology redesigns without disrupting network operations, and hence, eliminate unnecessary calls to SmartConnect repair phase.   
\item Finally, since RPL performs topology redesigns automatically, we need to figure out when a call to the SmartConnect repair phase is \emph{necessary}. We propose two empirical ways to do this:
\begin{enumerate}
\item Since we observe from Figure~\ref{fig:rpl-smartconnect} that RPL always recovered within 100 packets, we may wait for one 100-packets window, and if the QoS does not meet the target even after that, we shall invoke SmartConnect repair phase.
\item For $\alpha=0.5$, the minimum number of 20-packets windows required by RPL to change its PER estimate of a link from 1 to say, 0.01, assuming no packet drops on that link over that period, is $\lceil\frac{\ln 0.01}{\ln 0.5}\rceil=7$. Hence, wait for $7\times 20=140$ packets before triggering the SmartConnect repair phase.  
\end{enumerate}
\end{enumerate}

\subsection{Experiments with $k=1$}
\label{subsec:rpl-k1}
Next, we ask the question whether the redundancy in the underlying network topology for $k=2$ helps RPL performance, or whether RPL is flexible enough to deliver the same performance over the same time duration even on a sparser topology, say on a SmartConnect based network topology for $k=1$. To answer this question, we performed the following experiment.

\begin{figure}[t]
\centering
\includegraphics[scale=0.35]{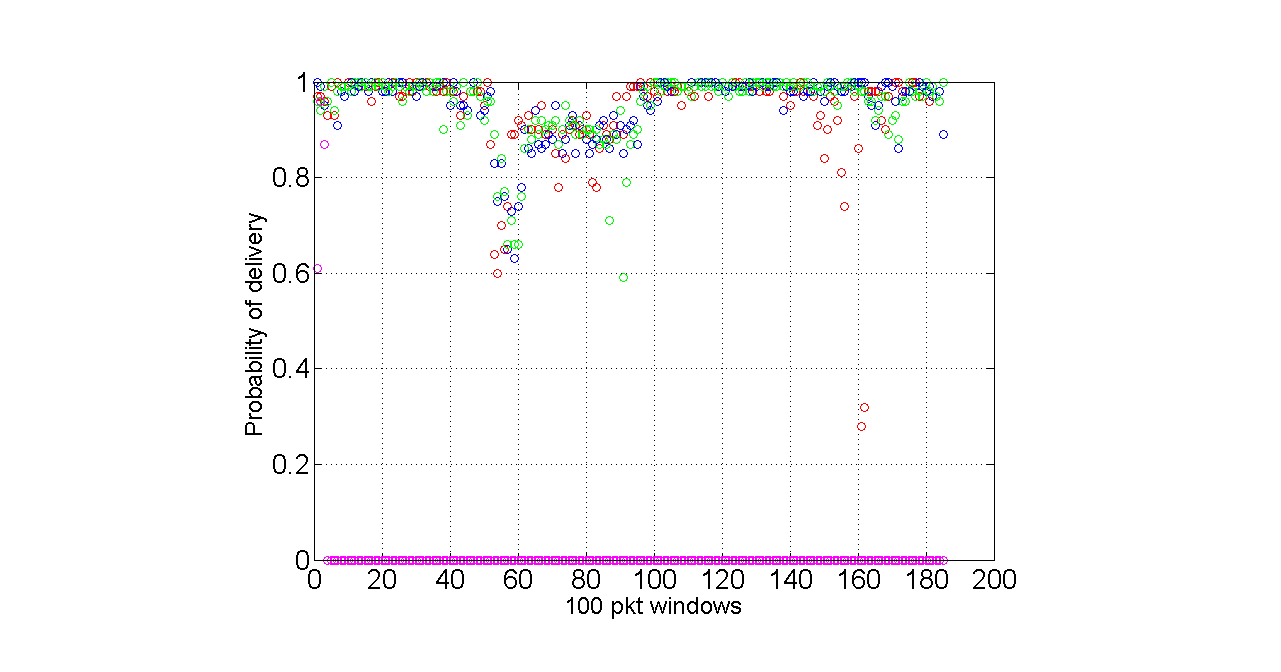}
\vspace{1mm}
\includegraphics[scale=0.35]{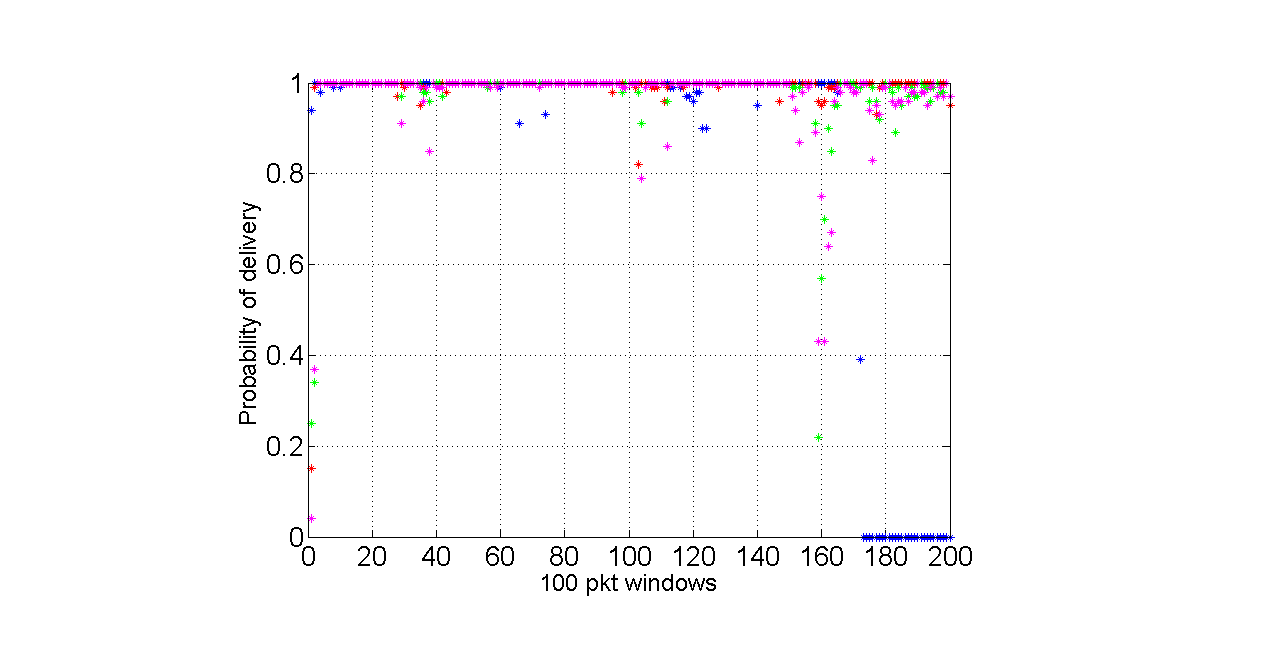}
\caption{Results of using RPL in conjunction with SmartConnect for $k=1$. Top panel: performance of static routing; Bottom panel: performance of RPL}
\label{fig:rpl-smartconnect2}
\vspace{-3mm}
\end{figure}

\gap
\noindent
\textbf{The experiment:}
\begin{enumerate}
\item Using SmartConnect, we designed a network for the same 4 sources as in Section~\ref{subsec:rpl-k2}, and $k=1$. The design parameters were the same as in the robustness experiments described in Section~\ref{sec:robustness}. The resulting topology is shown in Figure~\ref{fig:rpl-smartconnect-design2}.
\begin{figure}[t]
\centering
\includegraphics[scale=0.2]{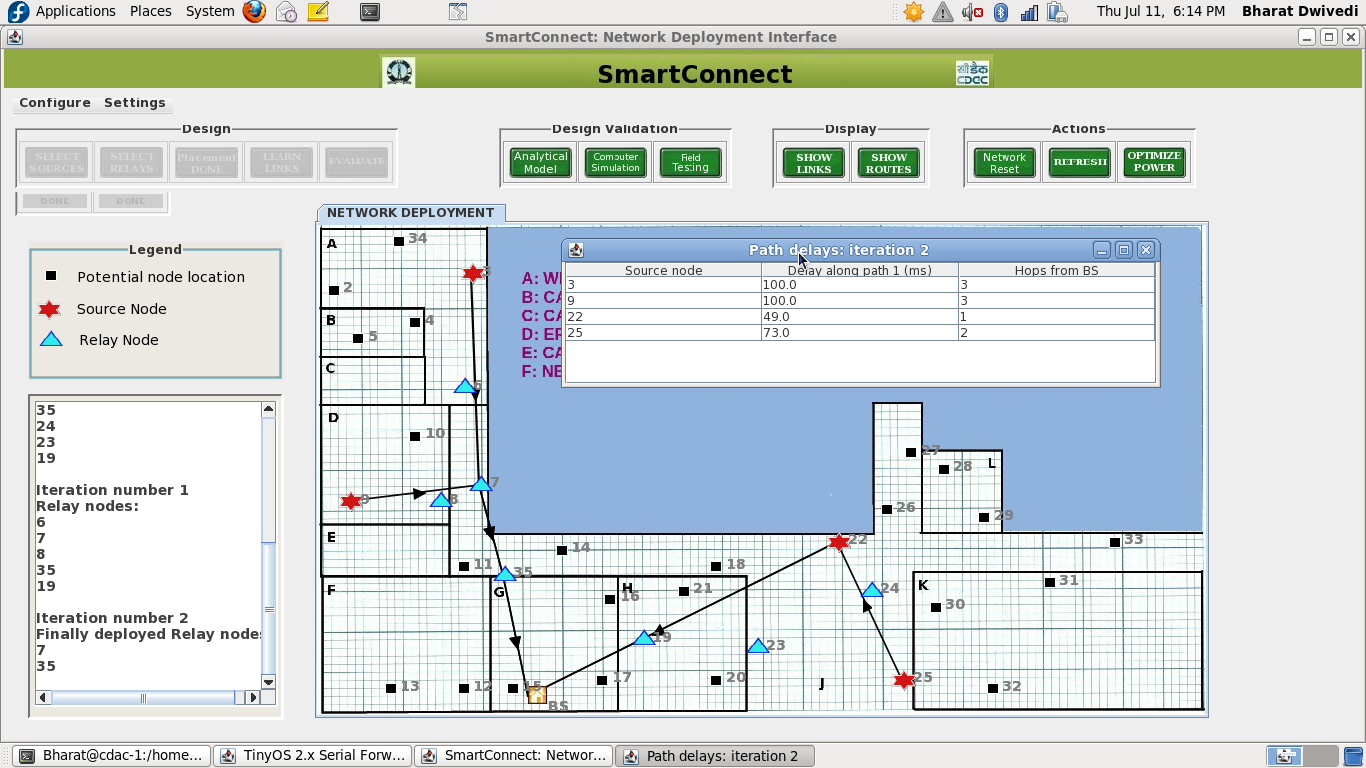}
\caption{Design obtained for $k=1$ using SmartConnect; RPL was run after deploying relays at the suggested locations}
\label{fig:rpl-smartconnect-design2}
\vspace{-3mm}
\end{figure}
\item The rest of the procedure was the same as in the previous experiment described in Section~\ref{subsec:rpl-k2}, except that no tracerouting was performed for SmartConnect since each source had only one static route to the sink.
\end{enumerate}
The results are summarized in Figure~\ref{fig:rpl-smartconnect2}.

\gap
\noindent
\textbf{Observations and Discussion:}
\begin{enumerate}
\item Comparing Figures~\ref{fig:rpl-smartconnect}, and \ref{fig:rpl-smartconnect2}, we see that
\begin{enumerate}
\item On the $k=2$ deployment, RPL achieved a probability of delivery close to 1 for a duration of 250 windows, i.e., in excess of 4 days, whereas on the $k=1$ deployment, after about 170 windows, i.e., 3 days, the probability of delivery for a source dropped to zero, and never recovered. Interestingly, the probability of delivery for that source in SmartConnect remained more than 80\% upto about 183 windows, indicating that \emph{a QoS-satisfying path existed, but RPL was unable to find it}. While it seems counter-intuitive at first glance, this situation can arise if a node in the current RPL routing tree \emph{does not have any alternate potential parent} to switch to when the link to its current preferred parent goes down. Clearly, such a situation is more likely in a sparse network deployment. 
\item Unlike in the $k=2$ case, in the $k=1$ case, we see more fluctuations in the RPL performance during the first 3 days, and it did not always recover within one window. 
\end{enumerate} 
\item From Figure~\ref{fig:rpl-smartconnect2}, SmartConnect probability of delivery for one of the sources (indicated by magenta color in the plot) dropped to zero after about 2 windows (50 minutes), and never recovered, whereas RPL continued to maintain a high $p_{\mathrm{del}}$ for that source. 
 
\end{enumerate}
To summarize, Observation 1 above clearly emphasizes the need for a redundant deployment, while Observation 2 once again demonstrates the advantage of RPL over SmartConnect static routing. Moreover, the observations suggest the counter-intuitive fact that \emph{neither RPL nor static routing is guaranteed to perform better than the other in all scenarios}; the performance depends on the underlying deployment. 

Overall, the experiments in Sections~\ref{subsec:rpl-k2} and \ref{subsec:rpl-k1} suggest that using RPL in conjunction with the robust network design obtained from SmartConnect may significantly enhance QoS performance of the network, and also eliminate unnecessary calls to the SmartConnect repair phase, thus causing minimal disruption to network operations. 
\section{Conclusion}
\label{sec:conclusion}

We have presented SmartConnect, a tool for assisting in designing and
deploying multihop relay networks for connecting wireless sensors with
a control center, for noncritical monitoring and control
applications. We described the core idea of field interactive
iterative design, and the associated procedures and algorithms.  The
SmartConnect system has been fully implemented and can be used for
network design in a variety of environments.

The core topology design algorithm, that SmartConnect currently uses,
assumes a light traffic model, so that at any time, with a high
probability just one packet is traversing the network. In our ongoing
work, we aim to extend SmartConnect to be able to design relay
networks for more general sensing loads. It may also be interesting to see how the network topologies designed using SmartConnect behave in conjunction with a dynamic routing protocol such as RPL \cite{phinney,rpl-rfc}. Finally, in this work, we have assumed that the nodes in the network are always awake. The problem of QoS aware network design for sleep-wake cycling nodes remains a topic of our future research.  

\section*{Acknowledgement} 
This work was supported by the Department of
Electronics and Information Technology (DeitY, Govt. of India) via the Automation Systems Technology (ASTEC) program, and by the Department of Science and
Technology (DST, Govt. of India) via a J.C.~Bose Fellowship. 
\vspace{-3mm}

\bibliographystyle{abbrv}
\bibliography{sigproc}  

\begin{thebibliography}{10}

\bibitem{sc-demo}
http://www.ece.iisc.ernet.in/dit-astec/smartconnect-demo.html.

\bibitem{senzaanalyzer}
www.e-senza.com/products/senzaanalyzer.

\bibitem{wireless-hart}
www.hartcomm.org/protocol/wihart/wireless\_technology.html.

\bibitem{ISA-100.11a}
www.isa.org/isa100.

\bibitem{dissemination-tep118}
www.tinyos.net/tinyos-2.x/doc/html/tep118.html.

\bibitem{vykon}
www.vykon.com.

\bibitem{abhijitmeth}
A.~Bhattacharya.
\newblock {N}ode {P}lacement and {T}opology {D}esign for {P}lanned {W}ireless
  {S}ensor {N}etworks.
\newblock {M}.{E} thesis, Indian Institute of Science, June 2010.

\bibitem{iwqos}
A.~Bhattacharya and A.~Kumar.
\newblock Delay {C}onstrained {O}ptimal {R}elay {P}lacement for {P}lanned
  {W}ireless {S}ensor {N}etworks.
\newblock In {\em 18th IEEE International Workshop on Quality of Service
  (IWQoS)}, 2010.

\bibitem{bhattacharya-techreport}
A.~Bhattacharya and A.~Kumar.
\newblock Qo{S} {A}ware and {S}urvivable {N}etwork {D}esign for {P}lanned
  {W}ireless {S}ensor {N}etworks.
\newblock Technical report, available at arxiv.org/pdf/1110.4746, 2011.

\bibitem{chen-terzis}
Y.~Chen and A.~Terzis.
\newblock {O}n the {I}mplications of the {L}og-normal {P}ath {L}oss {M}odel:
  {A}n {E}fficient {M}ethod to {D}eploy and {M}ove {S}ensor {M}otes.
\newblock In {\em Sensys}. ACM, 2011.

\bibitem{chipara}
O.~Chipara, G.~Hackmann, C.~Lu, W.~D. Smart, and G.-C. Roman.
\newblock {P}ractical {M}odeling and {P}rediction of {R}adio {C}overage of
  {I}ndoor {S}ensor {N}etworks.
\newblock In {\em IPSN}. ACM, 2010.

\bibitem{dawans}
S.~Dawans, S.~Duquennoy, and O.~Bonaventure.
\newblock On link estimation in dense {RPL} deployments.
\newblock In {\em 7th IEEE International Workshop on Practical Issues in
  Building Sensor Network Applications (SenseApp)}, pages 956--959, 2012.

\bibitem{huang-etal08deployment-zigbee-wsn}
Y.-K. Huang, P.-C. Hsiu, W.-N. Chu, K.-C. Hung, A.-C. Pang, T.-W. Kuo, M.~Di,
  and H.-W. Fang.
\newblock An integrated deployment tool for zigbee-based wireless sensor
  networks.
\newblock In {\em EEE/IFIP International Conference on Embedded and Ubiquitous
  Computing}, 2008.

\bibitem{IEEE}
IEEE.
\newblock {\em IEEE Standards Part 15.4: Wireless Medium Access Control (MAC)
  and Physical Layer (PHY) Specifications for Low-Rate Wireless Personal Area
  Networks (LR-WPANs)}.
\newblock New York, October 2003.

\bibitem{krause}
A.~Krause, C.~Guestrin, A.~Gupta, and J.~Kleinberg.
\newblock {N}ear-optimal {S}ensor {P}lacements: {M}aximizing {I}nformation
  while {M}inimizing {C}ommunication {C}ost.
\newblock In {\em IPSN}. ACM, 2006.

\bibitem{li-etal06planning-deployment-wsn}
J.~Li, Y.~Bai, J.~Ma, H.~Ji, and D.~Qian.
\newblock The architecture of planning and deployment platform for wireless
  sensor networks.
\newblock In {\em International Conference on Wireless Communicationss,
  Networking and Computing (WiCOM)}, 2006.

\bibitem{liu-kerpa}
T.~Liu and A.~E. Cerpa.
\newblock {F}oresee (4{C}): {W}ireless {L}ink {P}rediction using {L}ink
  {F}eatures.
\newblock In {\em IPSN}. ACM, 2011.

\bibitem{phinney}
T.~Phinney, P.~Thubert, and R.~Assimiti.
\newblock Rpl applicability in industrial networks:
  draft-phinney-roll-rpl-industrial-applicability-00.
\newblock Internet-Draft, October 2011.

\bibitem{raman-etal06link-(in)stability}
B.~Raman, K.~Chebrolu, N.~Madabhushi, D.~Y. Gokhale, P.~K. Valiveti, and
  D.~Jain.
\newblock Implications of link range and (in)stability on sensor network
  architecture.
\newblock In {\em WiNTECH}, September 2006.

\bibitem{ray09planning-analysis-wsn-tool}
A.~Ray.
\newblock Planning and {A}nalysis {T}ool for {L}arge {S}cale {D}eployment of
  {W}ireless {S}ensor network.
\newblock {\em International Journal of Next-Generation Networks (IJNGN)},
  1(1), 2009.

\bibitem{robinson-etal10deploying-nonuniform-propagation}
J.~Robinson, M.~Singh, R.~Swaminathan, and E.~Knightly.
\newblock Deploying mesh nodes under non-uniform propagation.
\newblock In {\em IEEE INFOCOM}, March 2012.

\bibitem{srinivasan-etal06RSSI-underappreciated}
K.~Srinivasan and P.~Levis.
\newblock Rssi is under appreciated.
\newblock In {\em Third Workshop on Embedded Networked Sensors (EmNets)}, 2006.

\bibitem{winet.srivatasava-etal12performance-analysis}
R.~Srivastava and A.~Kumar.
\newblock Performance analysis of beacon-less ieee 802.15.4 multi-hop networks.
\newblock In {\em Fourth International Conference on Communication Systems and
  Networks (COMSNETS)}, January 2012.

\bibitem{rpl-rfc}
T.~Winter, P.~Thubert, and R.~A. Team.
\newblock {RPL}: {I}pv6 routing protocol for low power and lossy networks,
  March 2012.

\end{thebibliography}
\end{document}